\documentclass{article}

\usepackage{amsmath}
\usepackage[ruled,vlined]{algorithm2e}
\usepackage{graphicx}
\usepackage{fullpage}

\begin{document}
\title{A Local Method for Detecting Communities}
\author{ James P. Bagrow$^1$ and Erik M. Bollt$^{2,1}$ \\
\small{\emph{$^1$  Department of Physics, Clarkson University, Potsdam, NY 13699-5820, USA.}} \\
\small{\emph{$^2$ Department of Math and Computer Science, Clarkson University, Potsdam, NY 13699-5815, USA.}}
}

\maketitle

\begin{abstract}
We propose a novel method of community detection that is computationally inexpensive and
possesses physical significance to a member of a social network. This method is unlike many divisive and agglomerative techniques and is local in the sense that a community can be detected within a network without requiring knowledge of the entire network.   A global application of this method is also introduced. Several artificial and real-world networks, including the famous Zachary Karate club, are analyzed.


\end{abstract}


\begin{section}{Introduction}
	It has been shown in the past that many interesting systems can be represented as networks composed of vertices and edges~\cite{StrogatzExploring, BarabasiAlbertStatistical,NewmanStructure,DorogovtsevMendesEvolution}.  Such systems include the Internet~\cite{FaloutsosInternet}, social and friendship networks~\cite{ScottSocialHandbook}, food webs~\cite{DunnWilliamsMartinezFood},  and citation networks~\cite{RednerHow, NewmanCollab}.  For example, a social network may represent people as vertices and edges linking vertices when those people are on a first-name basis.

	A topic of current interest in the area of networks has been the idea of communities and their detection.  A Community could be loosely described as a collection of vertices within a graph that are densely connected amongst themselves while being loosely connected to the rest 	of the graph~\cite{WF_Social, FLGC_Self,RCCVP_Defining}.   Many networks exhibit such a community structure and this motivates our work.  This description, however, is somewhat vague and open to interpretation.  This leads to the possibility that different techniques for detecting these communities may lead to slightly different yet equally valid results.  We emphasize this variation in Section \ref{ImpactAlpha}.
	
	Several techniques have been proposed to detect community structure inside of a network.  The recent and highly successful betweenness centrality algorithm due to Newman and Girvan~\cite{Newman_Mixing, GN_Community, NewmanLesMis} performs well within a variety of networks but it is costly to compute ($\mathcal{O}(n^2m)$ on a graph with $n$ vertices and $m$ edges)~\cite{NewmanLesMis}. More importantly, while betweenness centrality has been shown to be a useful quantity for detecting community structure, it is knowledge not usually attainable to a vertex \emph{within the graph}.

	In this paper we ask, if a person were to move to a new town, what actions would he or she take to see what community or communities they belong to?  Most community detection methods using hierarchical clustering fall within two categories: divisive and agglomerative \cite{ScottSocialHandbook, NewmanLesMis}.  Both forms, including those using betweenness and other methods, are global algorithms and don't represent feasible actions that a member of a network could undertake to identify the network's community structure.  The method proposed here may better represent actions that members of a network would undertake to identify their own communities.  
\end{section}

\begin{section}{The Algorithm}\label{algorithm}
	The proposed algorithm consists of an $l$-shell spreading outward from a starting vertex.  As the starting vertex's nearest neighbors and next nearest neighbors, etc. are visited by the $l$-shell, two quantities are computed: the emerging degree and total emerging degree.  The emerging degree of a vertex is defined as  the number of edges that connect that vertex to vertices the $l$-shell has not already visited as it expanded from the previous $l-1, l-2, ...$ -shells.  Note that edges between vertices within the same $l$-shell do not contribute to the emerging degree.

Let us define the following notation for the emerging degree and total emerging degree:
	\begin{align}
		k_i^e(j)  = & \mbox{ emerging degree of vertex $i$ from a} \nonumber \\ 
		                  & \mbox{ shell started at vertex $j$.} \nonumber \\
		K_j^l = & \mbox{ total emerging degree of a shell of} \nonumber \\ 
				              & \mbox{ depth $l$ starting from vertex $j$.} 
		\label{Equation1}
	\end{align}

	The total emerging degree of an $l$-shell is then the sum of the emerging degrees of all vertices on the leading edge of the $l$-shell.  This can also be thought of as the total number of \emph{emerging edges} from that $l$-shell \cite{CostaHub}.   We see that the total emerging degree at depth $l$ is not necessarily the number of vertices at depth $l+1$.  At depth 0, the total emerging degree is just the degree of the starting vertex. At depth $l$, it is the total number of edges from vertices at depth $l$ connected to vertices at depth $>l$.
	
	It follows from Eqn. \ref{Equation1} that:
	\begin{align}
		K_j^0  &=  k_j \nonumber \\
		K_j^l  &=  \sum_{ i  \in  S_j^l} k_i^e(j)
	\end{align} 
	where $S_j^l$ is the leading edge of the $l$-shell, that is, the set of all vertices exactly $l$ steps away from vertex $j$.

	In addition, let us define the \emph{change in total emerging degree}:
	\begin{equation}
		\Delta K_j^l = \frac{K_j^l}{K_j^{l-1}} 
	\end{equation}
	for a shell at depth $l$ starting from vertex $j$.  

	
	The algorithm works by expanding an $l$-shell outward from some starting vertex $j$ and comparing the change in total emerging degree to some threshold $\alpha$.  When:
	\begin{align}
		\Delta K_j^l &< \alpha
	\end{align}
	the $l$-shell ceases to grow and all vertices covered by shells of a depth $\leq l$ are listed as members of vertex $j$'s community.  

	We describe our algorithm roughly as follows.  For a starting vertex $j$:

\begin{enumerate}
\itemindent=5mm
	\itemsep=1pt
\item Start an $l$-shell, $l=0$, at vertex $j$ (add $j$ to the list of community members) and compute $K_j^0$. 

\item Spread the $l$-shell, $l=1$, add the neighbors of $j$ to the list, and compute $K_j^1$

\item Compute $\Delta K_j^1$.  If $\Delta K_j^1 < \alpha$, then a community has been found.  Stop the algorithm.

\item Else repeat from step 2 for the next $l$-shell, until $\alpha$ is crossed or the entire connected component is added to the community list.
\end{enumerate}


	See Algorithm \ref{localAlg.alg} for more exact pseudocode.\\

	Since there tend to be many interconnections within a community, so definied, as an $l$-shell grows outward from some starting vertex within a community, the total emerging degree will tend to increase.  See Section \ref{Motivating} for more discussion of an idealized graph model with community structure.  When the $l$-shell reaches the ``border" of the community, the number of emerging edges will decrease 	sharply.  This is because, at this point, the only emerging edges are those connecting the community to the rest of the graph 		which are, by definition, less in number than those within the community.  

	By introducing a single parameter, $\alpha$, and monitoring $\Delta K_j^l$, the $l$-shell's growth can be stopped when it has 	covered the community.  It is this fact that allows for the starting vertex to detect its community locally:  at the last depth before 	$\alpha$ is crossed, it does not matter where the emerging edges lead.  See Section \ref{selectlocalresults} for results using our purely 	local method.

	Our method is not perfect, however, and it is possible for the $l$-shell to ``spill over" the community it is detecting.  This is dependent on how the starting vertex is situated within the graph: if it is closer (or equally close) to some non-community vertex or vertices than to some community vertices, the $l$-shell may spread along two or more communities at the same time.  To alleviate this effect, one can run the 	algorithm $N$ times, using each vertex as a starting vertex, and then achieve a \emph{group consensus} as to which vertices belong to which communities.  This idea is discussed in Section \ref{globalinfo}.

	\begin{algorithm}
		  \SetLine
		  \dontprintsemicolon
		  $s \in V$; \emph{// $s$ is the starting vertex}\;
		  $K_s^{d-1}  \longleftarrow$ 1;\;
		  $Q \longleftarrow$ empty queue; \emph{// search queue}\;
		  $C \longleftarrow$ empty queue; \emph{// community queue}\;
		  enqueue $s \longrightarrow Q$;\;
		  $K_s^d \longleftarrow$ $emerging\left(Q, C, G(V,E)\right)$;\;
		  $\Delta K_s^d \longleftarrow \frac{K_s^d}{K_s^{d-1}}$; \;
		  \BlankLine
		  \BlankLine
		  \While{$\ \Delta K_s^d > \alpha$}{
		    $K_s^{d-1} \longleftarrow K_s^d$;\;
		    \ForEach{$q \in Q$}{
		      dequeue $q \longleftarrow Q$;\;
		      enqueue $q \longrightarrow C$;\;
		      enqueue $neighbors(q) \longrightarrow Q$;\;
		      }
		    $ K_s^d \longleftarrow$ $emerging\left(Q, C, G(V,E)\right)$;\;
		    $\Delta K_s^d \longleftarrow \frac{K_s^d}{K_s^{d-1}}$; \;
		    }
		  \caption{Local algorithm to determine a starting vertex's community~\protect\footnotemark.  Note that $emerging$ is a function of the $l$-shell and the graph that returns the total emerging degree.}
		  \label{localAlg.alg}
	\end{algorithm}

	The idea of having an expanding $l$-shell encompass a community is not in itself new here.  The hub-based algorithm in~\cite{CostaHub} expands 	multiple $l$-shells simultaneously from the $n$ vertices of highest degree (the hubs) until all vertices are within an $l$-shell.  While 	computationally inexpensive, this method has the following drawback:  the number of communities detected is arbitrarily pre-assigned and the algorithm neglects the 	possibility of having two hubs within the same community.  In addition, it requires knowledge of the entire graph; it is a global algorithm, not local.

\footnotetext{Note that this algorithm differs slightly from the one outlined in the text of Section \ref{algorithm}.  In the pseudocode we have essentially chosen a $K_j^{-1} = 1$ to seed the algorithm.  This has an impact on $\alpha$:  if $\alpha$ is larger than the degree of the starting vertex, then the $l$-shell's growth will terminate immediately and the result will be a singleton community.  Throughout the paper, we have used the algorithm described in the text:  $\Delta K_j^1$ is the first value compared to $\alpha$ and requires gathering both $K_j^0$ (which is just the degree of vertex $j$) and $K_j^1$.  When starting the algorithm in this way, the $l$-shell will always spread to the immediate neighbors of the starting vertex before the change in total emerging degree is compared to $\alpha$.  This results in all runs listing the neighbors of the starting vertex as member's of that starting vertex's community, regardless of $\alpha$.  This only impacts the results for large $\alpha$ (which usually lead to a final result of $N$ singleton communities anyway) and has no effect on the results presented here.}

	\begin{subsection}{Select Local Results}
	\label{selectlocalresults}
		We have applied Algorithm \ref{localAlg.alg} to the Zachary Karate club as shown in Figure \ref{KarateLocals}.   Figure \ref{ActualKarateNewman} shows the actual split the club underwent. We complete two runs, one starting from vertex 17 and another from vertex 24.  Five vertices (3, 9, 14, 20, and 32) were claimed by both runs as members of that starting vertex's community.  Please note that this graph and all subsequent graphs and dendrograms were drawn using \cite{GraphvizCite}.  
		
		We interpret our results as follows.  The five vertices that are listed as members of both starting vertex's communities tend to fall on the ``border" between the two groups.  This makes sense to us since each vertex is linked roughly equally to both communities.  One can imagine these five members had the most difficult choice to make when the club split.  Far from being an unwanted result,  this overlap could be used to predict vertices that may be more likely to switch communities in the future (in an evolving network) or which vertices are least isolated within a single community.  
	\end{subsection}

	\begin{figure}
		\begin{minipage}[t]{0.475\linewidth} 
			\begin{center}
				\includegraphics[width=.8\textwidth]{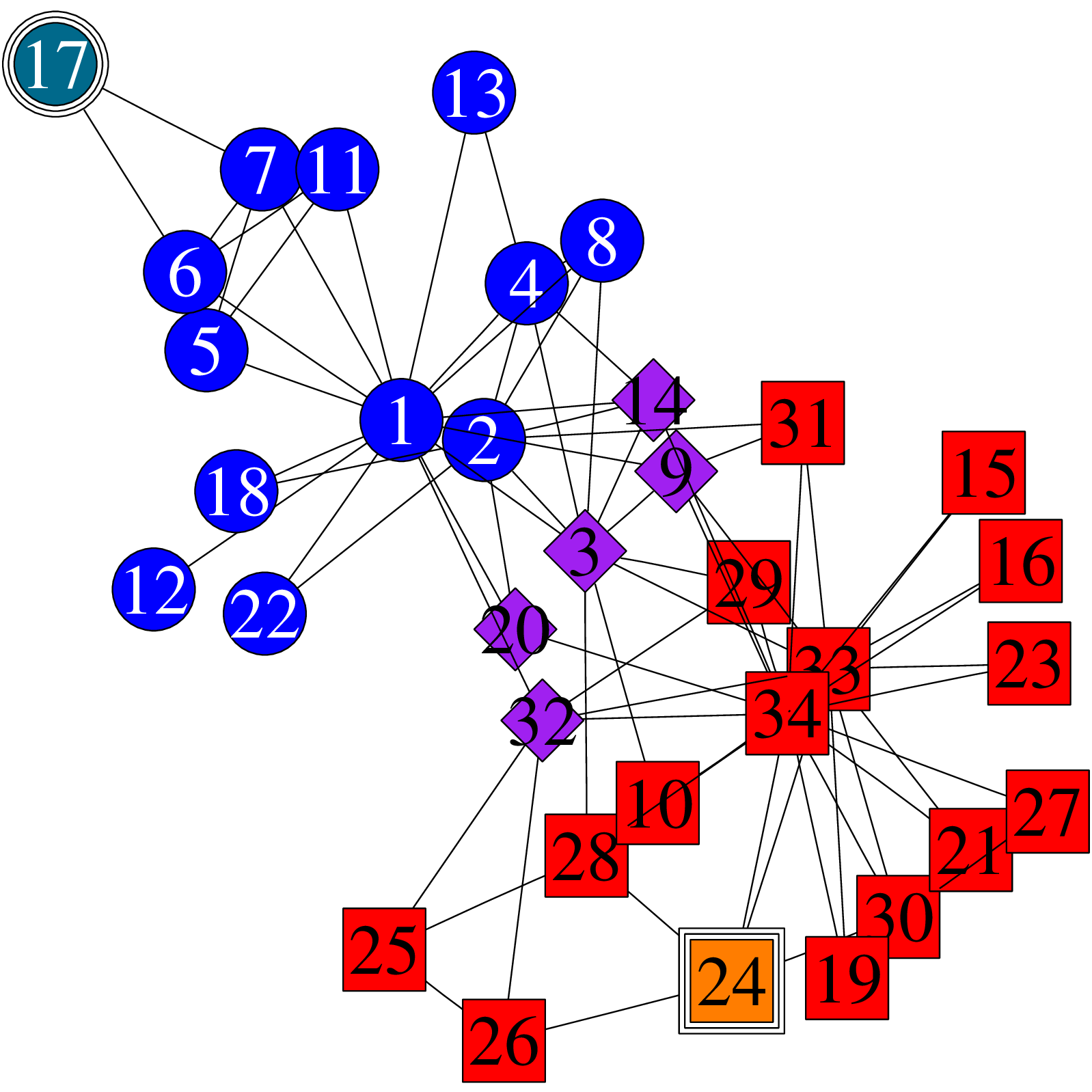}
				\caption{Two local results on the Zachary Karate Club with $\alpha=1.9$.  Boxes and diamonds represent the output of Algorithm \protect\ref{localAlg.alg} when starting from vertex 24, while circles  and diamonds represent the output when starting from vertex 17.  }
				\label{KarateLocals}
			\end{center}
		\end{minipage}
		\hspace{0.5cm} 
		\begin{minipage}[t]{0.475\linewidth}
			\begin{center}
				\includegraphics[width=\textwidth]{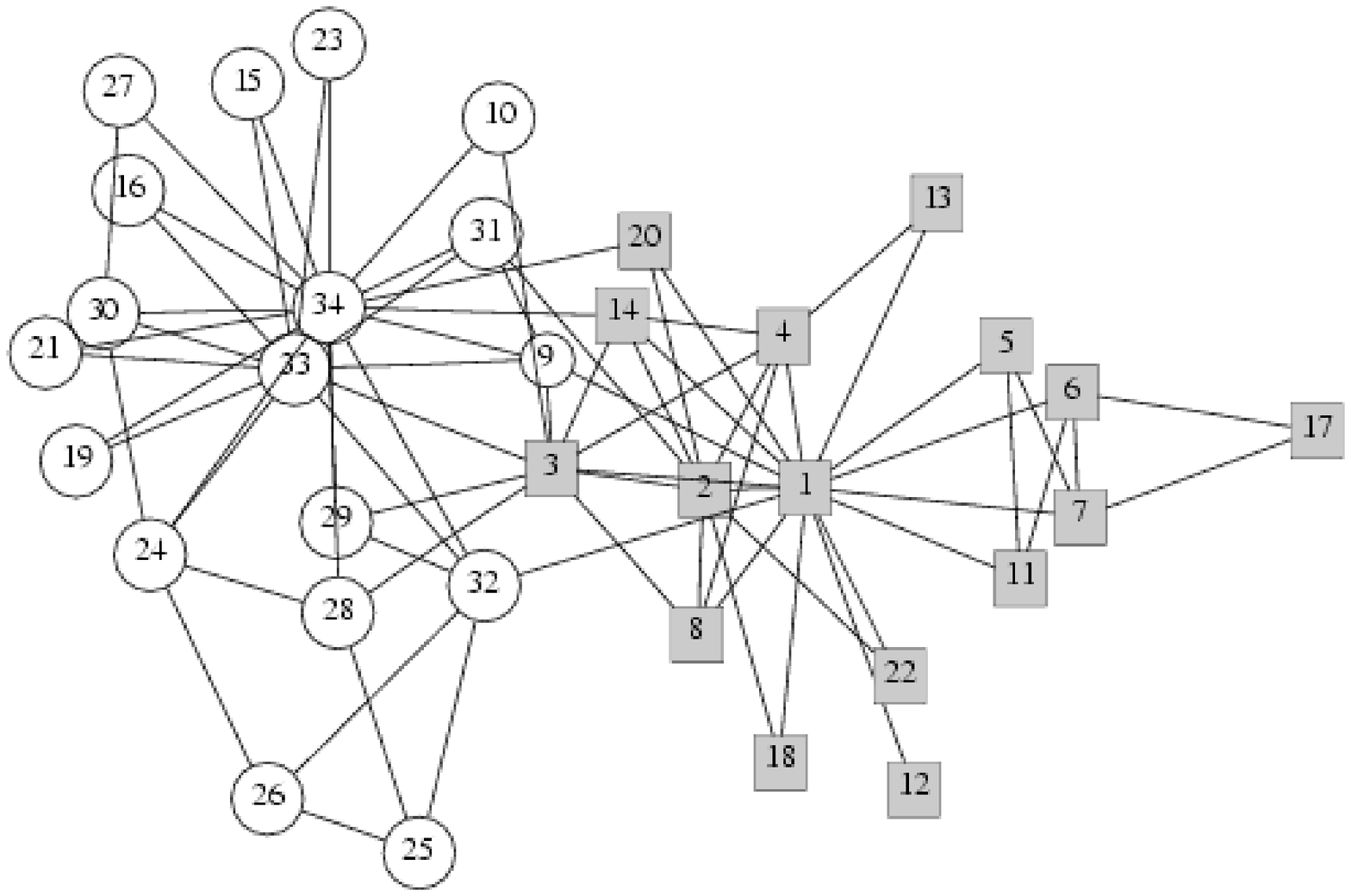}
				\caption{Actual breakdown of the Karate Club.~\cite{NewmanSite}}
				\label{ActualKarateNewman}
			\end{center}
		\end{minipage}
	\end{figure}
	
	\begin{subsection}{Obtaining Global Information}\label{globalinfo}
	
Algorithm \ref{localAlg.alg} is a method for a single vertex to determine something about its community membership. It seems reasonable that, by surveying all the locally-determined membership listings, one should be able to generate an idea of the global structure of the network. Here we propose a simple method using a \emph{membership matrix} to obtain such a picture and to overcome membership overlap (discussed in Section \ref{selectlocalresults}) when determining a ``consensus" partitioning of the network.

For any given starting vertex $j$, Algorithm \ref{localAlg.alg}  can return a vector, $\mathbf{v}_j$ of size $N$, where the $i$th component is 1 if vertex $i$ is a member of the starting vertex's community and 0 otherwise. These vectors can be assembled to form a $N$ x $N$ membership matrix, 
		\begin{equation}
		M = \left( \mathbf{v}_1 | \mathbf{v}_2 | ... | \mathbf{v}_N \right)^t
		\end{equation}
where the $j$th row contains the results from using vertex $j$ as the algorithm's starting point.  This allows for a good way to visualize the resultant data when starting the algorithm from multiple vertices.

We define a \emph{Distance} between rows $i$ and $j$ of the membership matrix as the total number of differences between their components:
		\begin{equation}
			\mbox{Distance}(i, j) = n - \sum_{k=1}^n \delta\left( M_{ik}, M_{jk} \right)
		\end{equation}
		where $\delta\left( M_{ik}, M_{jk} \right)=1$ if $M_{ik}=M_{jk}$ and $0$ otherwise.

		Now we perform a simple sorting algorithm on $M$.  For the $i$th row:
\begin{enumerate}
\itemindent=5mm
\itemsep=1pt
\item Find Distance$(i,j)$ for all rows $j>i$.

\item Pick the row that is the smallest Distance to row $i$ (call it row $k$) and interchange it with row $i+1$.  This requires swapping rows $i+1$ and $k$ and swapping columns $i+1$ and $k$.  Columns are swapped because a row interchange is equivalent to a renumbering of the involved vertices, so that new numbering must be kept consistent throughout $M$.

\item Repeat for row $i+1$.\\
\end{enumerate}	

Unfortunately, the sorting step can be computationally expensive.  Finding Distance$(i,j)$ costs $\mathcal{O}(N)$.  When the sorting algorithm begins at the first row, there are $N-1$ Distances to find, so the cost of the first sort is $\mathcal{O}(N(N-1))\sim \mathcal{O}(N^2$).  This is then repeated for the next row, costing $\mathcal{O}(N(N-2))\sim \mathcal{O}(N^2$), and so on for each additional row.  Since there are $N$ rows, the total cost is:
		\begin{equation}
		\sum_{i=1}^N N\left(N-i \right) = N \left( N^2 - \frac{1}{2}N\left(N+1\right) \right) = \mathcal{O}(N^3)
		\end{equation}
		
		The result of this sorting/renumbering is a membership matrix that is much more indicative of structure.  Specifically, we have a sorted membership matrix,
		\begin{equation}
		\tilde{M} = P^t M P
		\end{equation}
		where $P$ is a permutation matrix effectively resulting from the above.  Well-separated communities appear as blocks along the diagonal, and imperfections within the blocks can indicate substructure (See Figures  \ref{KarateMMbeforeSort} and \ref{KarateMMafterSort}). 
		
			\begin{figure}
				\begin{minipage}[t]{0.475\linewidth} 
						\includegraphics[width=\textwidth]{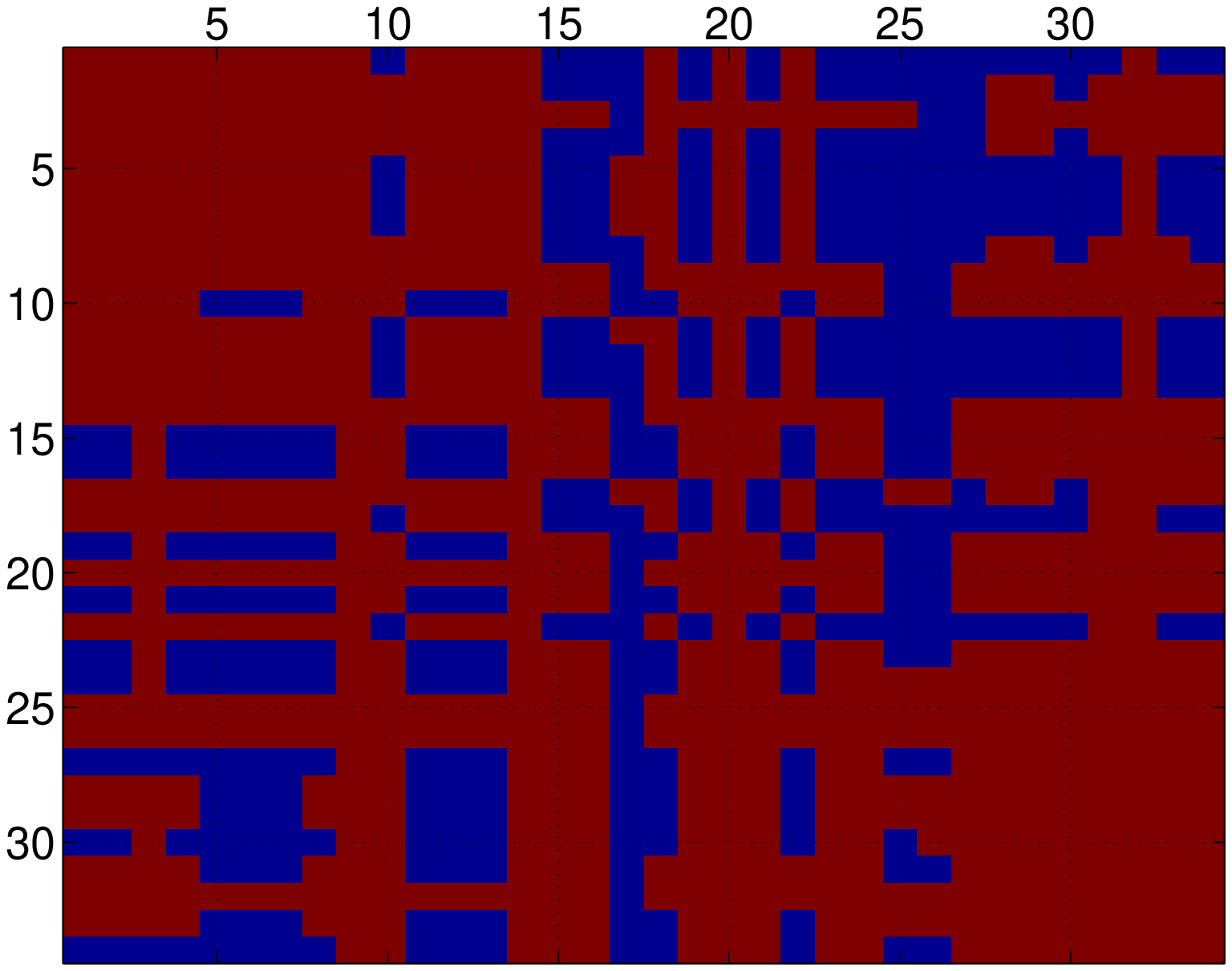}
						\caption{Membership matrix $M$ for the Zachary Karate Club before sorting, with $\alpha=1.2$.  Red boxes  indicate a 	value of 1, blue 0. }
						\label{KarateMMbeforeSort}
				\end{minipage}
				\hspace{0.5cm} 
				\begin{minipage}[t]{0.475\linewidth}
						\includegraphics[width=\textwidth]{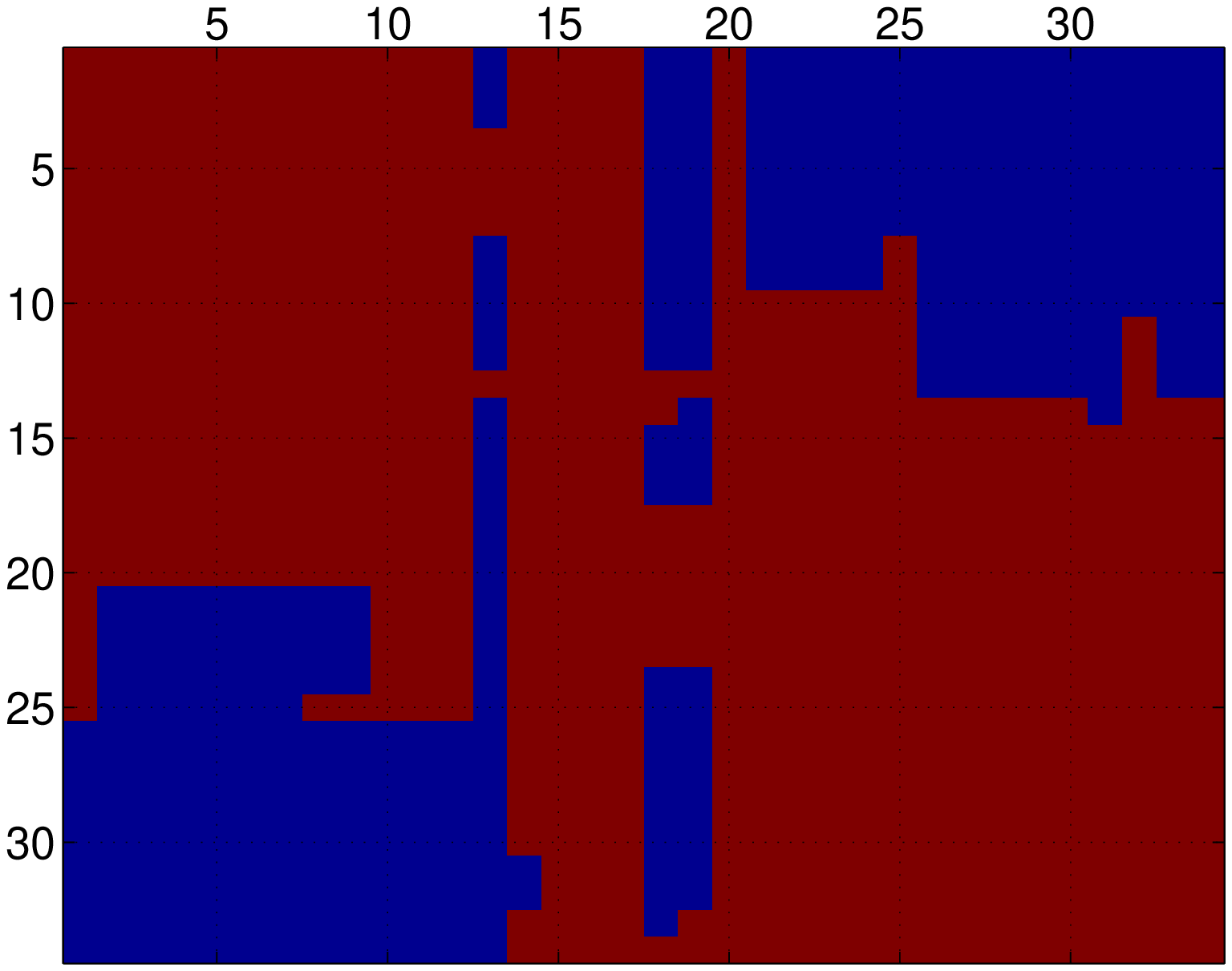}
						\caption{Sorted membership matrix $\tilde{M}$ for the Karate Club , with $\alpha=1.2$. }
						\label{KarateMMafterSort}
				\end{minipage}
			\end{figure}
	\end{subsection}

	\begin{subsection}{Finding a hierarchy of sub-communities} \label{SubComms}

Sorting the membership matrix already provides a useful means of visualizing the results of all the different runs of the local algorithm, but this is not enough to determine how any present sub-communities relate to larger communities.  Therefore, here we introduce a further operation to apply to $\tilde{M}$ to generate a dendrogram of  the community structure.
		  
		For row $i$, we compute a cumulative row distance, CD$_i$:
		 \begin{align}
		\mbox{CD}_1 &= 0 \nonumber \\
		\mbox{CD}_i &= \mbox{Distance}(i,i-1) + \mbox{CD}_{i-1} \nonumber \\
			 &= \sum_{j=2}^i  \mbox{Distance}(j-1,j) 
		\end{align}
		 
Plotting the row number $i$ versus the cumulative distance CD$_i$  will yield a collection of points of increasing value falling into discrete bands that indicate the members of each community.  Note that the row number $i$ is the new sorted number $i$ for that vertex: one needs to keep track of all the individual sorting operations to maintain the original number of that vertex, that is, through the permutations $P$.  This step of finding these cumulative distances costs $\sim \mathcal{O}(N^2)$ operations.  See Figures \ref{Ideal1RD},  \ref{Ideal2RD},  \ref{Ideal3RD},  and \ref{KarateRD},  for plots of examples of these cumulative row distances for various networks.  These plots are useful for visualization but are not strictly necessary to get the sub-community hierarchy.

Finally, to yield a dendrogram of the community structure, the following operation is performed:  

\begin{enumerate}
	\itemindent=5mm
	\itemsep=1pt
	\item $d \longleftarrow1$.
	\item Compute Distance${(i-1,i)}$ for all $i=2 .. n$. 
	\item Choose the smallest Distance (often zero for identical rows) and call it $D_{min}$.
	\item $C_d \longleftarrow$ empty queue. 	 \emph{// clustering queue}
	\item enqueue  1st  vertex $\longrightarrow C_d$.
	\item FOR i = 2..n : 
	\begin{enumerate}
		\itemindent=10mm
		\itemsep=1pt
		\item IF Distance${(i-1,i )}>$ $D_{min}$: 
		\begin{enumerate}
			\itemindent=15mm
			\itemsep=1pt
			\item $d \longleftarrow d+1$.
			\item $C_d \longleftarrow$ empty queue. 	
		\end{enumerate}
		\item enqueue $i$th vertex $\longrightarrow C_d$.
	\end{enumerate}
	\item Repeat from 3 for next smallest Distance until all Distances have been used.
\end{enumerate}

Essentially, we are moving down the rows of $\tilde{M}$ and grouping together all the vertices whose corresponding rows are closer together than $D_{min}$ until we arrive at a row that is farther away than $D_{min}$.  Then we start a new group and begin grouping the subsequent vertices together until we  \emph{again} find a row that is farther away than $D_{min}$, and so forth.    This is then repeated using the next smallest Distance as $D_{min}$.  This has a course-graining effect: as we use larger Distances for $D_{min}$, farther vertices will start grouping together.  

Grouping the rows of $\tilde{M}$ in this way is equivalent to grouping the vertices of the graph together into a sub-community hierarchy.  This is also similar in form to many agglomerative techniques, with the row distances of $\tilde{M}$ used as a similarity measure.  These groupings can then be used to generate a dendrogram of the sub-community structure if we assume that each vertex is a singleton before we started grouping and that after the largest Distance is used,  all vertices are grouped together.   See Figures \ref{Ideal1Dendo}, \ref{Ideal2Dendo}, \ref{Ideal3Dendo}, \ref{KarateDendo}, \ref{BooksDendo}, and \ref{LesMisDendo} for such dendrograms.
\end{subsection}
	
\begin{subsection}{The impact of $\alpha$}\label{ImpactAlpha}
	
The algorithm is based on a single parameter, $\alpha$, which controls when to stop the spread of the $l$-shell.  When $\alpha=0$, the $l$-shell will never stop until the entire connected component has been visited.  As $\alpha$ increases in size, $l$-shells will tend to stop growing sooner, until eventually they do not spread beyond the starting vertex and the final result will be $N$ singleton communities.  This is guaranteed to happen when $\alpha > k_{max}$, where $k_{max}$ is the largest degree in the network.   
	
	The impact of varying $\alpha$ is readily apparent in Figures \ref{EvolveMMlow} and \ref{EvolveMMhigh}.  In Figure \ref{EvolveMMlow}, the smaller $\alpha$ allowed the $l$-shells to spread farther: many $l$-shells starting from vertices close to the main partition (the two edges of highest betweenness) have spread to the entire network.  In Figure \ref{EvolveMMhigh}, the larger value of $\alpha$ truncated the $l$-shells before they had a chance to spread beyond the sub-communities of the starting vertices. 
	
	In contrasting Figures \ref{EvolveMMlow} and \ref{EvolveMMhigh}, we emphasize that how one defines a communitiy through an algorithm bears on the specific results.  It is our contention that there is not a single true answer for a community partition.  We hold that the flexibility of a parameter like $\alpha$ to allow for various levels of community courseness is in fact quite natural: the result is in the eye of the beholder - of the specific algorithm.  In any case, as can be seen by our examples in Section \ref{RealWorldNets}, intermediate values of $\alpha$ lead to community partitions which agree well with many found in the literature, and we think that they make good sense in light of our model interpretation as described in the next section.  
	
	One can think of $\alpha$ as a measure of the  ``friendliness" of the starting vertex, to use a social network analogy.  When $\alpha$ is small ($ \alpha \ll 1$), the $l$-shells will spread to much of the network.  This can indicate vertices that are more likely to include other vertices in their respective communities or, in a social network, people who are more welcoming of their neighbors.   Similarly, when $\alpha$ is large ($\alpha \gg 1$), the $l$-shells will stop growing immediately.  This can be indicative of vertices that are unlikely to include other vertices in their community, or hermit-like people who are unwilling to accept even their immediate neighbors into their communities, instead preferring to remain a singleton.  In this sense, $\alpha$ can be thought of as an inverse measure of friendliness or social acceptance.  
	
		\begin{figure}
			\begin{minipage}[t]{0.475\linewidth} 
					\includegraphics[width=\textwidth]{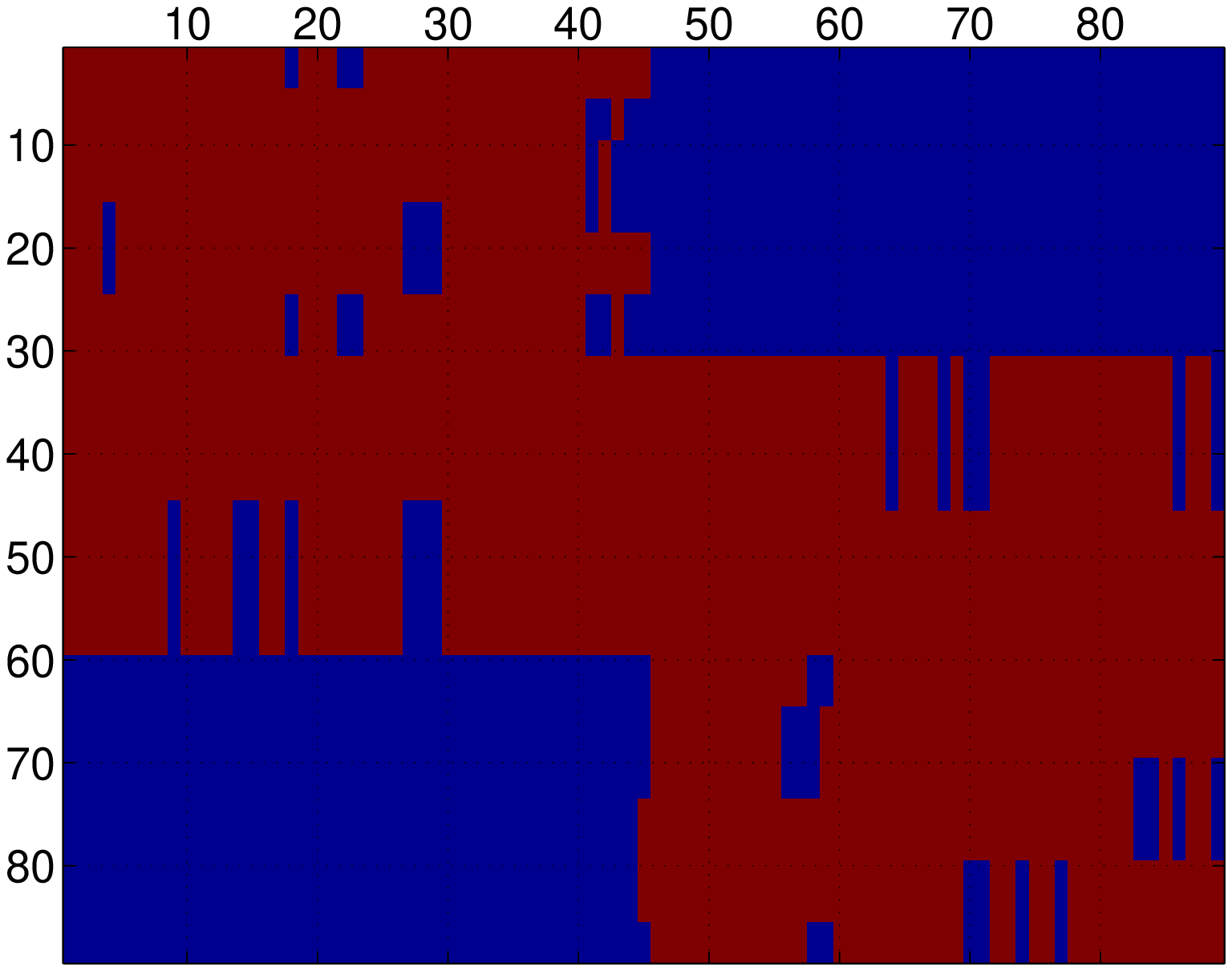}
					\caption{Membership matrix for Ideal graph 3 after sorting, with  $\alpha=0.25$.  Note the increased number of "spilled" vertices in the the middle rows, as compared with Figure \protect\ref{Ideal3MM}. }
					\label{EvolveMMlow}
			\end{minipage}
			\hspace{0.5cm} 
			\begin{minipage}[t]{0.475\linewidth}
				\begin{center}
					\includegraphics[width=\textwidth]{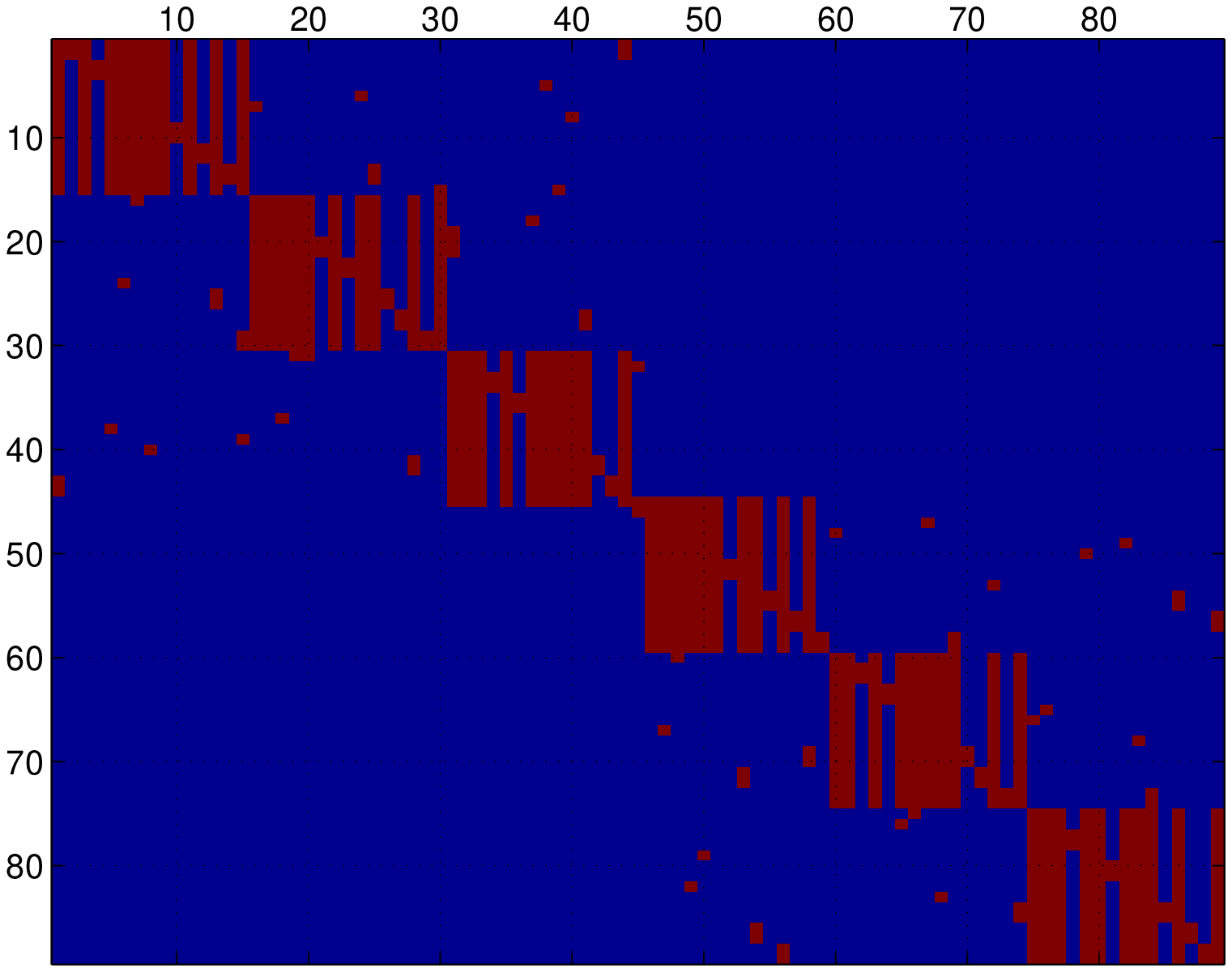}
					\caption{Membership matrix for Ideal graph 3 after sorting, with $\alpha=3$.  The larger value of $\alpha$ stops the $l$-shells sooner, allowing for the smaller sub-communities to become evident instead of the main partitioning.   }
					\label{EvolveMMhigh}
				\end{center}
			\end{minipage}
		\end{figure}
	
	\end{subsection}
\end{section}

\begin{section}{Motivating Examples}\label{Motivating}

	We propose the following idealized models for a network with simple community structure, and we test our algorithm first in these cases.  Our models demonstrate the high-density intraconnections and low-density interconnections.  We define an idealized community of size $N$ as a complete subgraph ($k_j=\langle k \rangle = N-1$), with one or more additional edges linking it to one or more additional ideal communities.
	
	These networks represent the extreme fulfillment of the idea of a community.  Each community has the maximum number of internal links possible while having close to the minimum number of external links.  This results in the number of emerging edges dropping off very sharply at the border of each subgraph, leading to nearly identical results when $\alpha$ and the starting vertices are varied.
	
	Several configurations of these ideal networks are analyzed (Figures \ref{Ideal1graph}, \ref{Ideal2graph}, and \ref{Ideal3graph},).  These networks also provide a means of visualizing, understanding, and interpreting how the membership matrix may look for a given community structure.  In addition, these networks contain single vertices situated between the subgraphs.  Since these vertices are equally connected to multiple communities, the results from Algorithm \ref{localAlg.alg} starting from these vertices will contain all the subgraphs that the vertices are linked to.  This is evident in the rows of the membership matrix that overlap two or more blocks.
	
	Through these models, we can better understand how to interpret the performance of our algorithm;  we believe these models in fact make suitable benchmarks for other community partitioning algorithms found in the literature.  One can easily assemble a graph with a given community structure, apply a community partitioning algorithm to it, and compare the results of that algorithm with the structure created when the graph was assembled. 
			
	In addition, it is useful to note that these networks require little or no sorting of their membership matrices.  This is because the vertices are already numbered consecutively: vertices 1 through $i$ are community 1, vertices $i+1$ through $j$ are community 2, etc.  This is, of course, a contrived result  and cannot be expected in general.

	\begin{figure}
		\begin{minipage}[t]{0.475\linewidth} 
			\begin{center}
				\includegraphics[width=.6\textwidth]{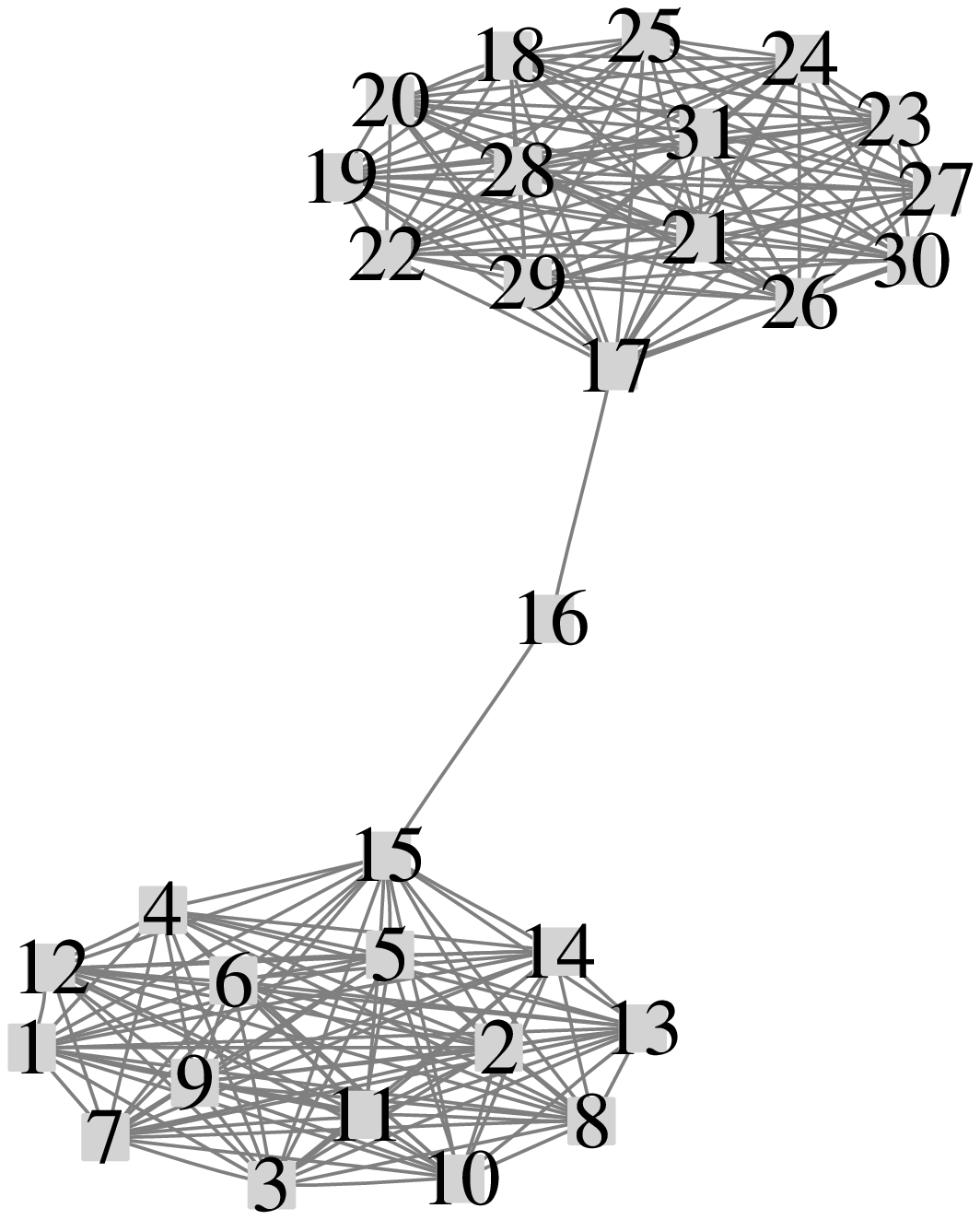}
				\caption{Ideal graph 1: Two complete subgraphs of size 15 bridged by a common vertex.}
				\label{Ideal1graph}
			\end{center}
		\end{minipage}
		\hspace{0.5cm} 
		\begin{minipage}[t]{0.475\linewidth}
			\begin{center}
				\includegraphics[width=\textwidth]{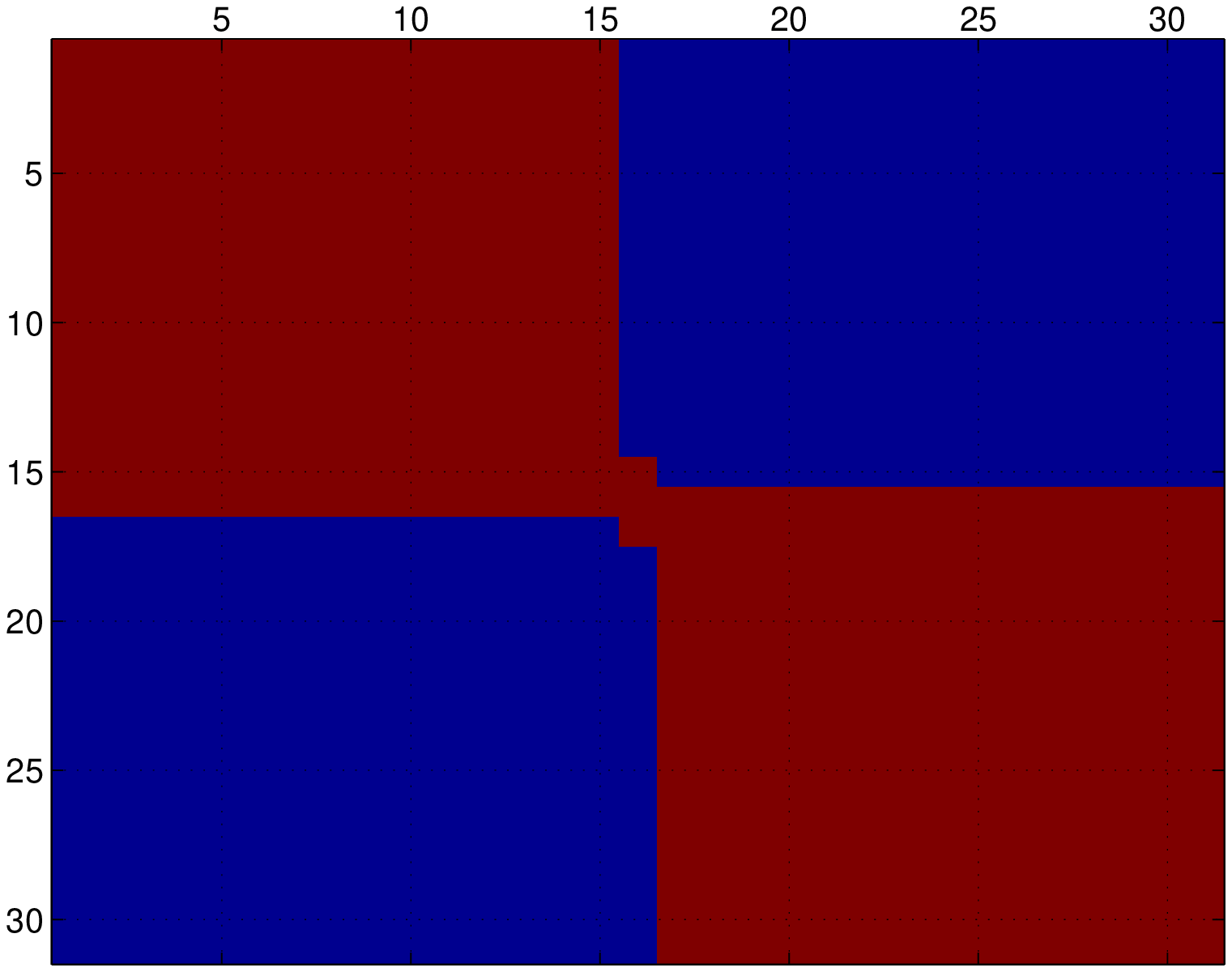}
				\caption{Membership matrix for Ideal graph 1.  Note that no sorting was required.  $\alpha=1$, but for these idealized model graphs, the results are identical for a wide range of $\alpha$ values.} \label{Ideal1MM}
			\end{center}
		\end{minipage}
		\begin{minipage}[b]{0.475\linewidth}
			\begin{center}
				\includegraphics[width=\textwidth]{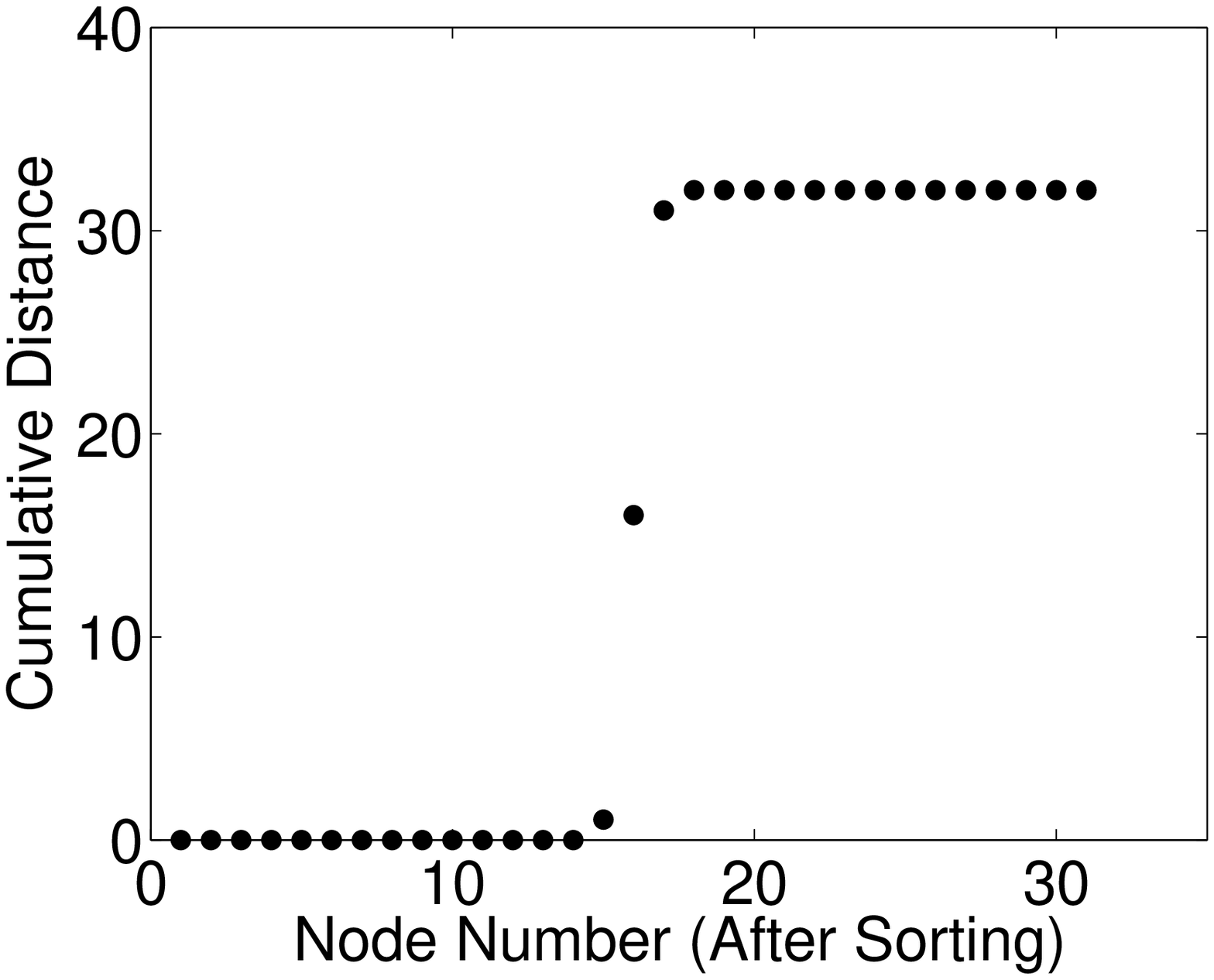}
				\caption{Cumulative row distances for Ideal graph 1.  Generated using the membership matrix shown in Figure \ref{Ideal1MM}. }
				\label{Ideal1RD}
			\end{center}
		\end{minipage}
		\hspace{0.5cm} 
		\begin{minipage}[b]{0.475\linewidth} 
			\begin{center}
				\includegraphics[width=\textwidth]{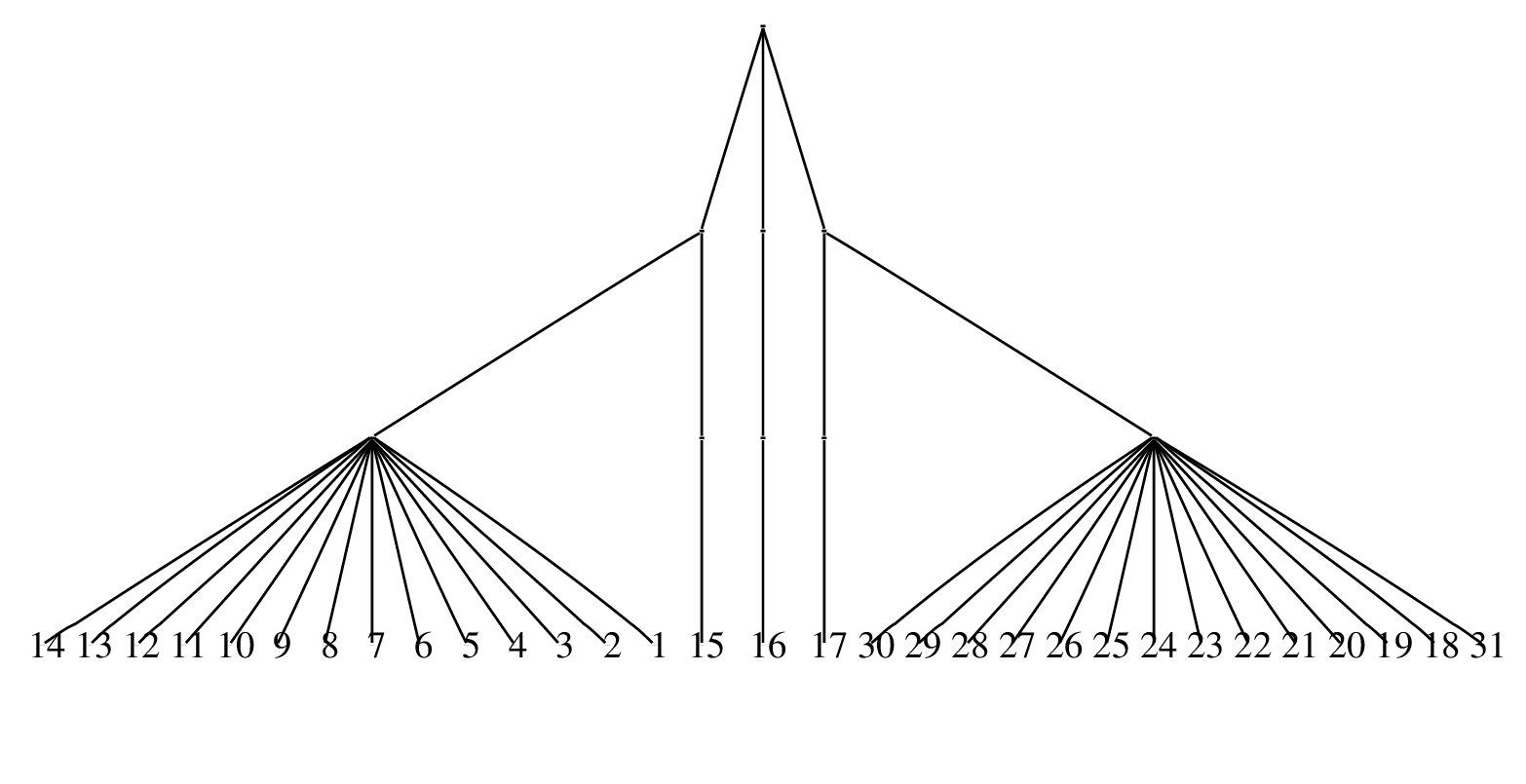}
				\caption{Dendrogram for Ideal graph 1.  Notice that central member vertex 16 is idealized here as a community unto itself, which in light of the form of the graph in Figure \ref{Ideal1graph}, simply means that 16 is central between two communities.  Notice also the special placement of members 15 and 17.}
				\label{Ideal1Dendo}
			\end{center}
		\end{minipage}
	\end{figure}
	
	\begin{figure}
		\begin{minipage}[t]{0.475\linewidth} 
			\begin{center}
				\includegraphics[width=.8\textwidth]{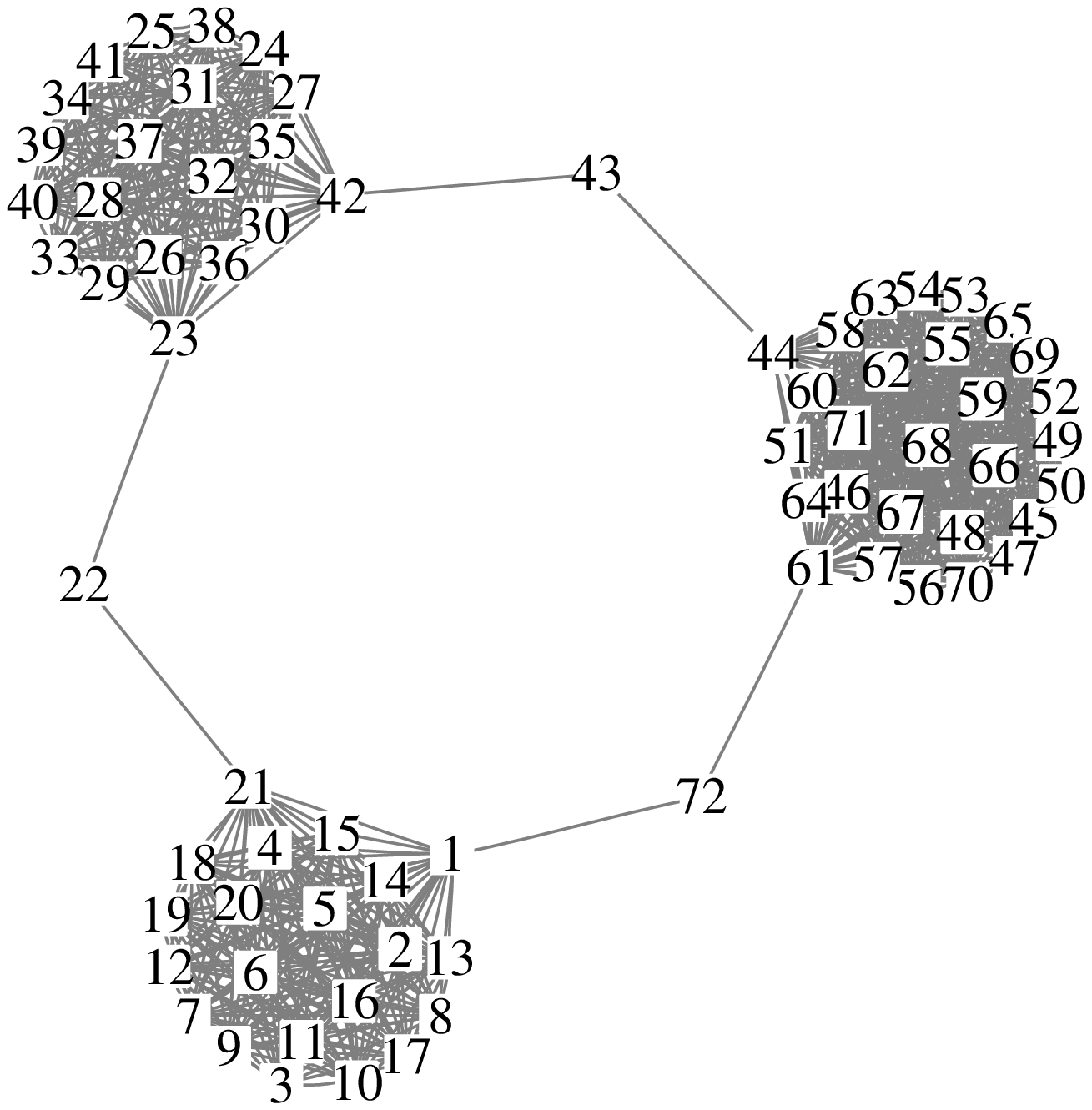}
				\caption{Ideal graph 2: Three complete subgraphs, one larger than the others.}
				\label{Ideal2graph}
			\end{center}
		\end{minipage}
		\hspace{0.5cm} 
		\begin{minipage}[t]{0.475\linewidth}
			\begin{center}
				\includegraphics[width=\textwidth]{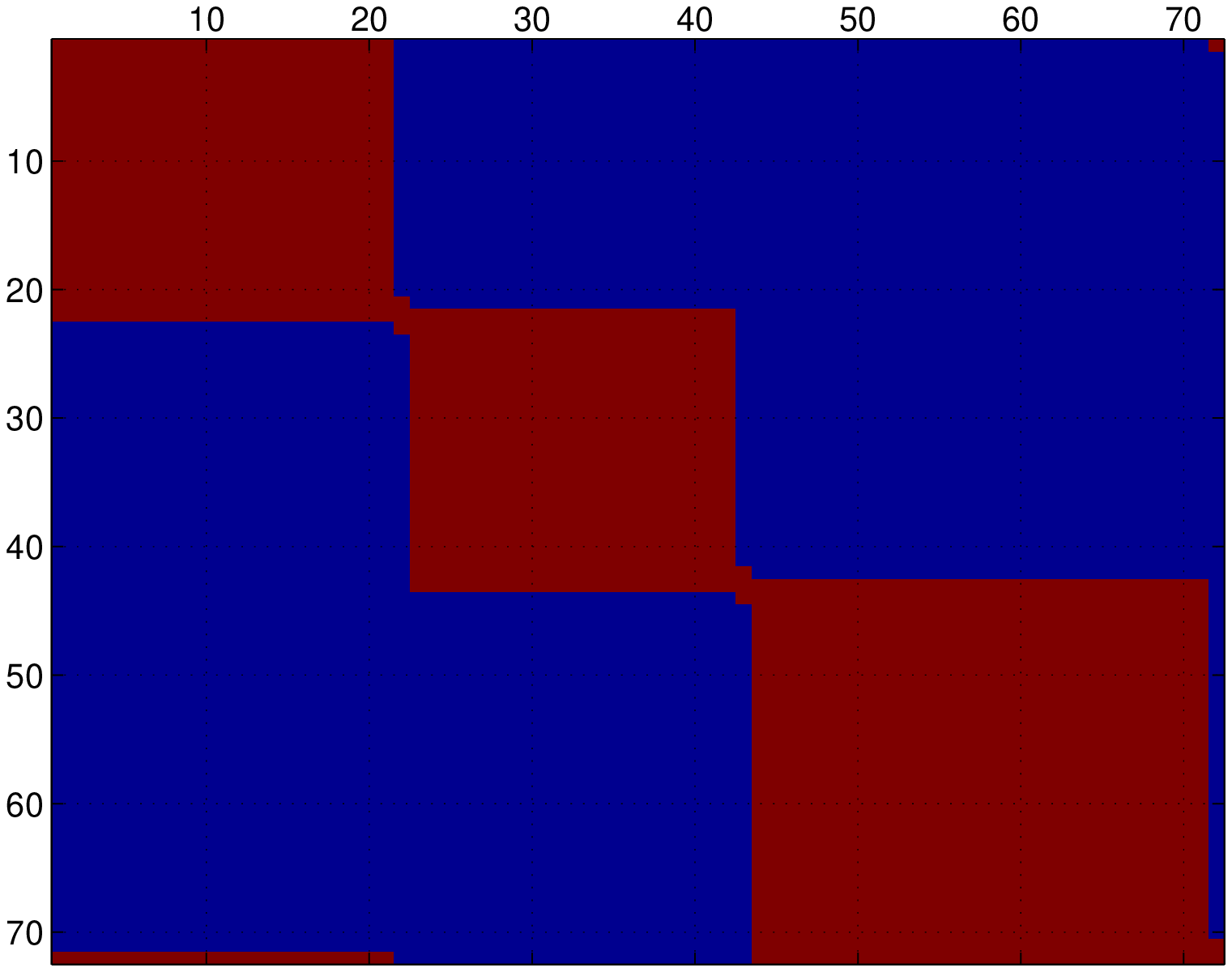}
				\caption{Membership matrix for Ideal graph 2.}
				\label{Ideal2MM}
			\end{center}
		\end{minipage}

		\begin{minipage}[b]{0.475\linewidth}
			\begin{center}
				\includegraphics[width=\textwidth]{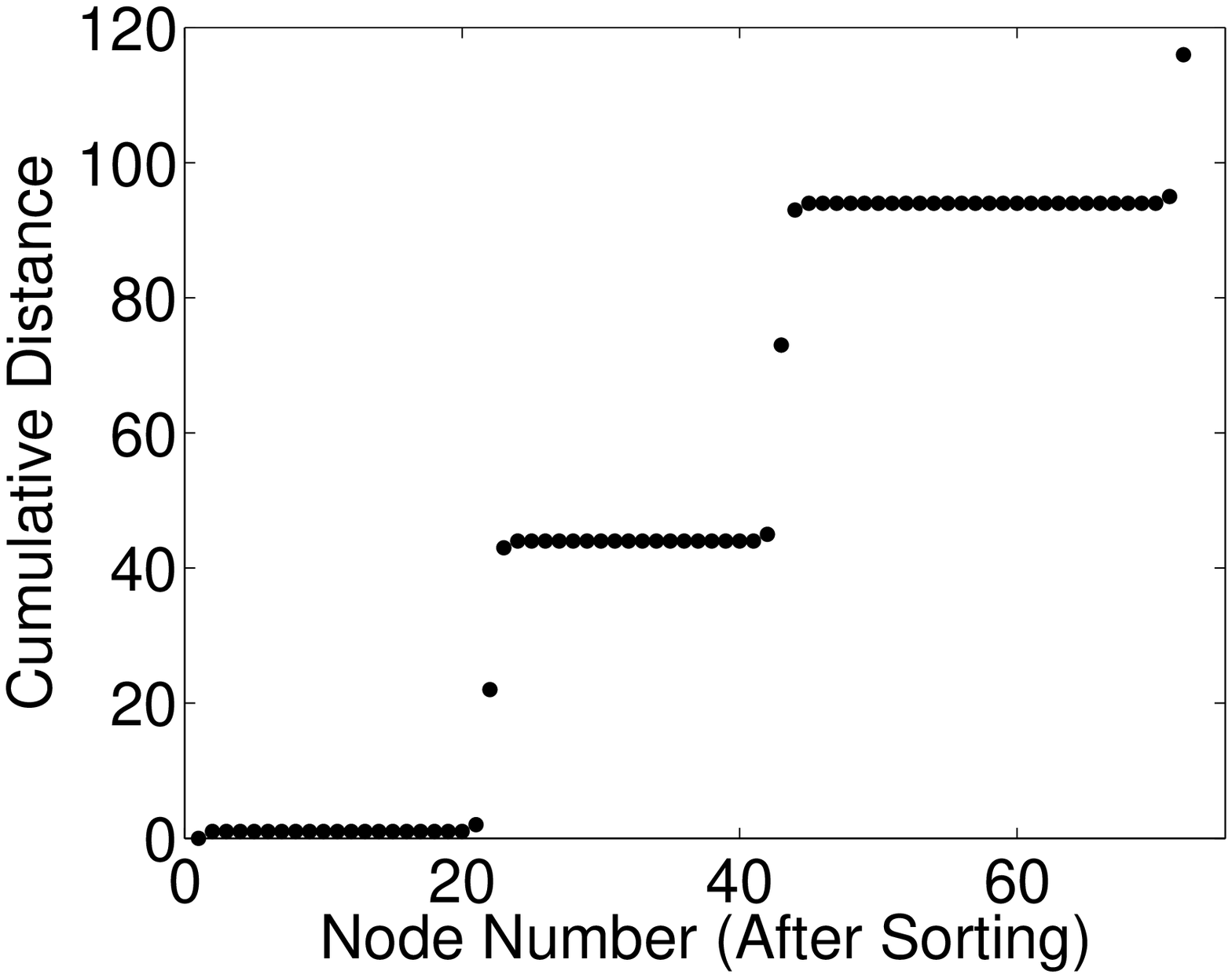}
				\caption{Row Distances for Ideal graph 2.  Generated using the membership matrix shown in Figure \ref{Ideal2MM}.}
				\label{Ideal2RD}
			\end{center}
		\end{minipage}
		\hspace{0.5cm} 
		\begin{minipage}[b]{0.475\linewidth} 
			\begin{center}
				\includegraphics[width=\textwidth]{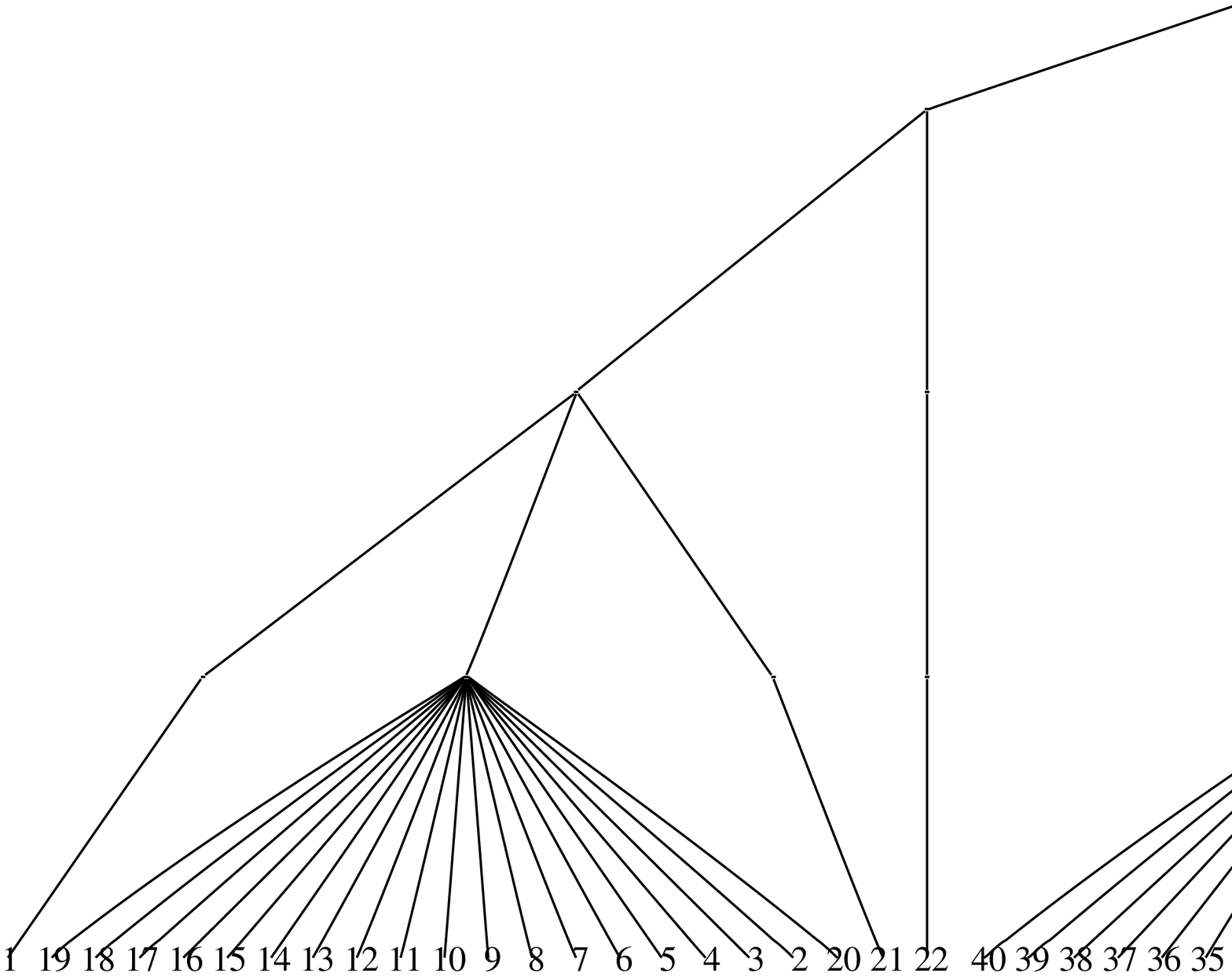}
				\caption{Dendrogram for Ideal graph 2.  Again, as in Figure \ref{Ideal1Dendo}, notice the results for central members 22, 43, 72, and also the boundary members 1, 21, 23, 42, 44, and 61. }
				\label{Ideal2Dendo}
			\end{center}
		\end{minipage}
	\end{figure}
	
	\begin{figure}
		\begin{minipage}[b]{0.475\linewidth} 
			\begin{center}
				\includegraphics[width=0.4\textwidth]{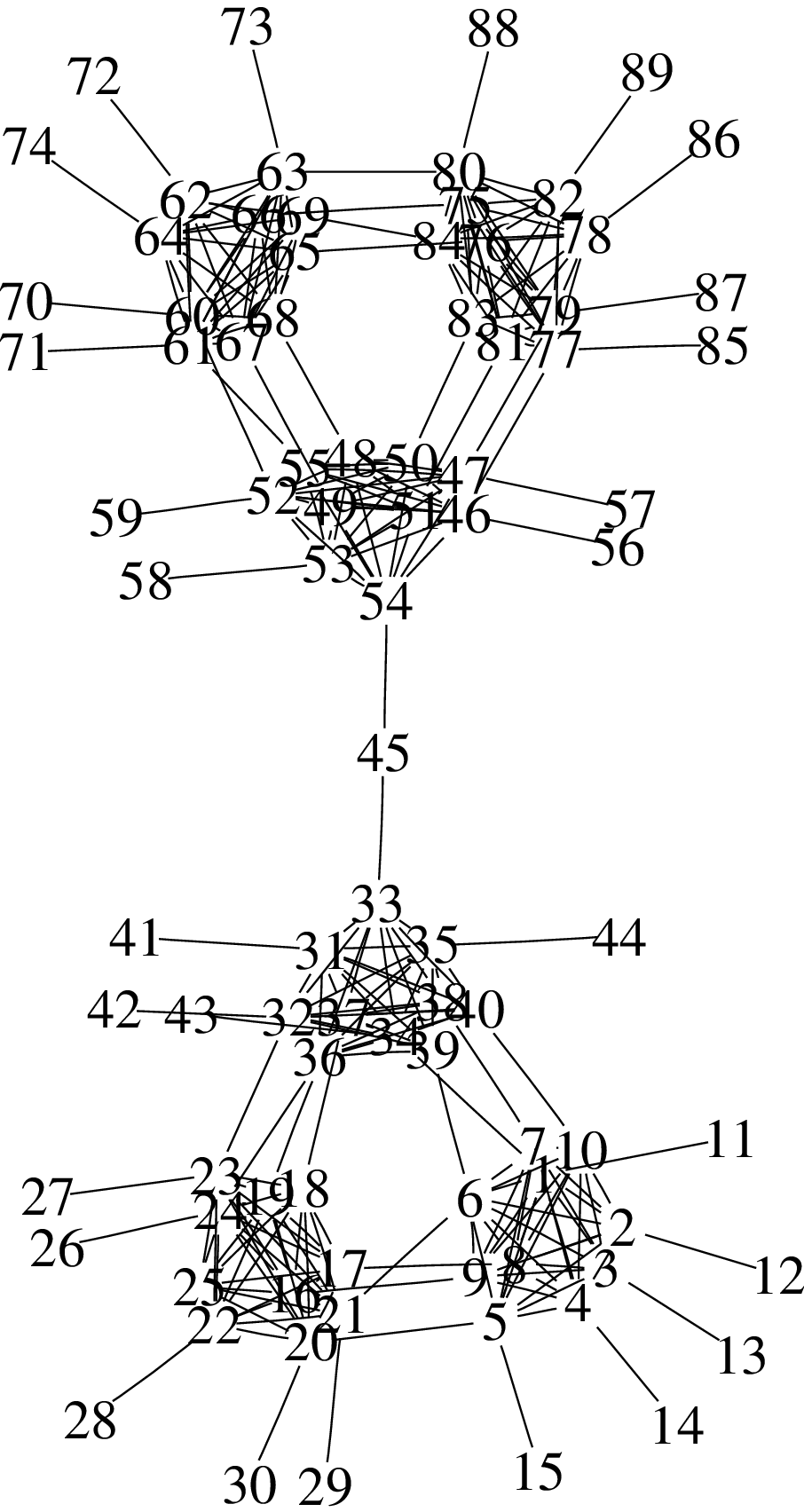}
				\caption{Ideal graph 3:  Three complete subgraphs joined together by multiple edges with singletons attached bridged by a single vertex to another group of similar (but not identical) structure.}
				\label{Ideal3graph}
			\end{center}
		\end{minipage}
		\hspace{0.5cm} 
		\begin{minipage}[b]{0.475\linewidth}
			\begin{center}
				\includegraphics[width=\textwidth]{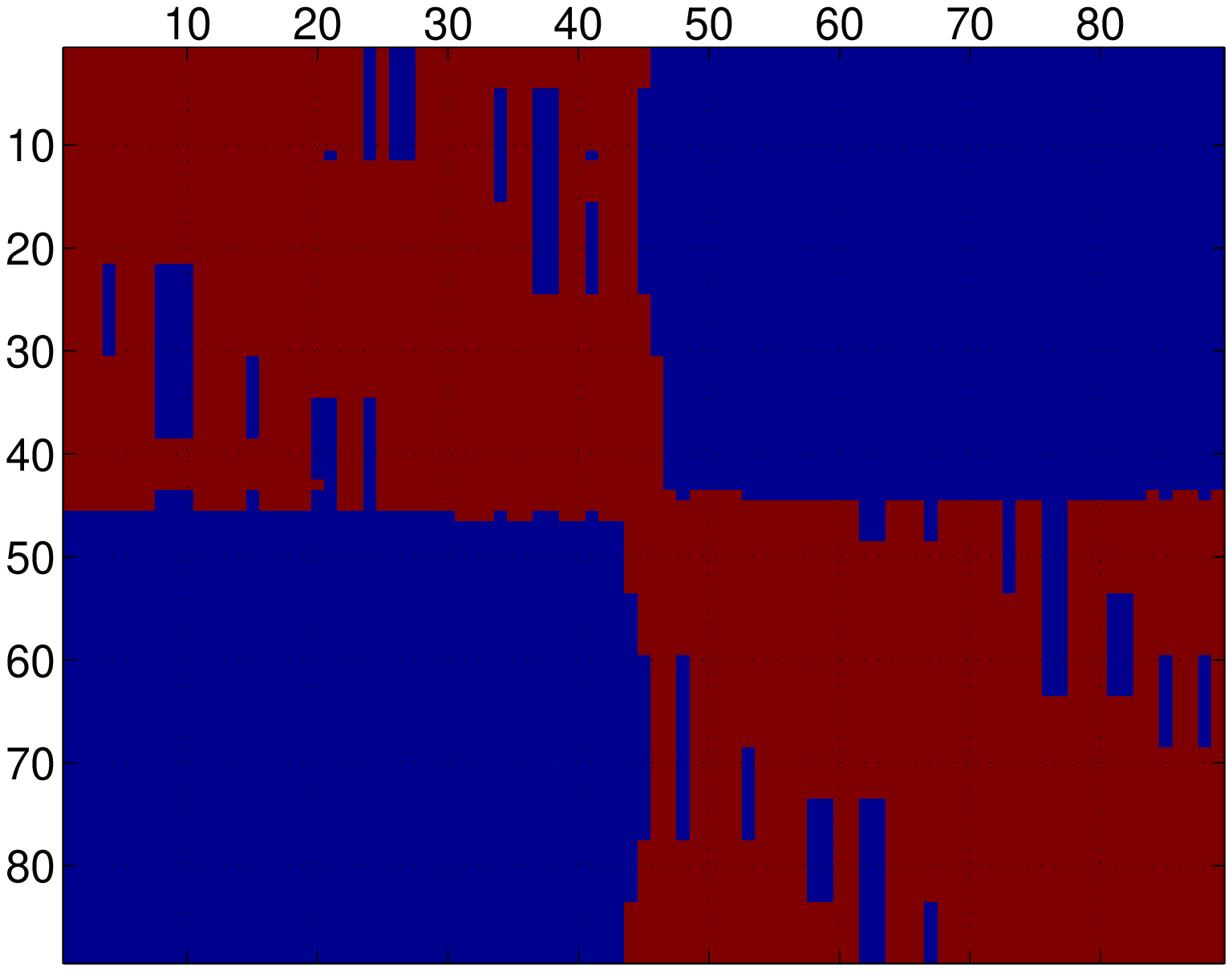}
				\caption{Membership matrix for Ideal graph 3, $\alpha=0.87$.}
				\label{Ideal3MM}
			\end{center}
		\end{minipage}
		\begin{minipage}[b]{0.475\linewidth}
			\begin{center}
				\includegraphics[width=\textwidth]{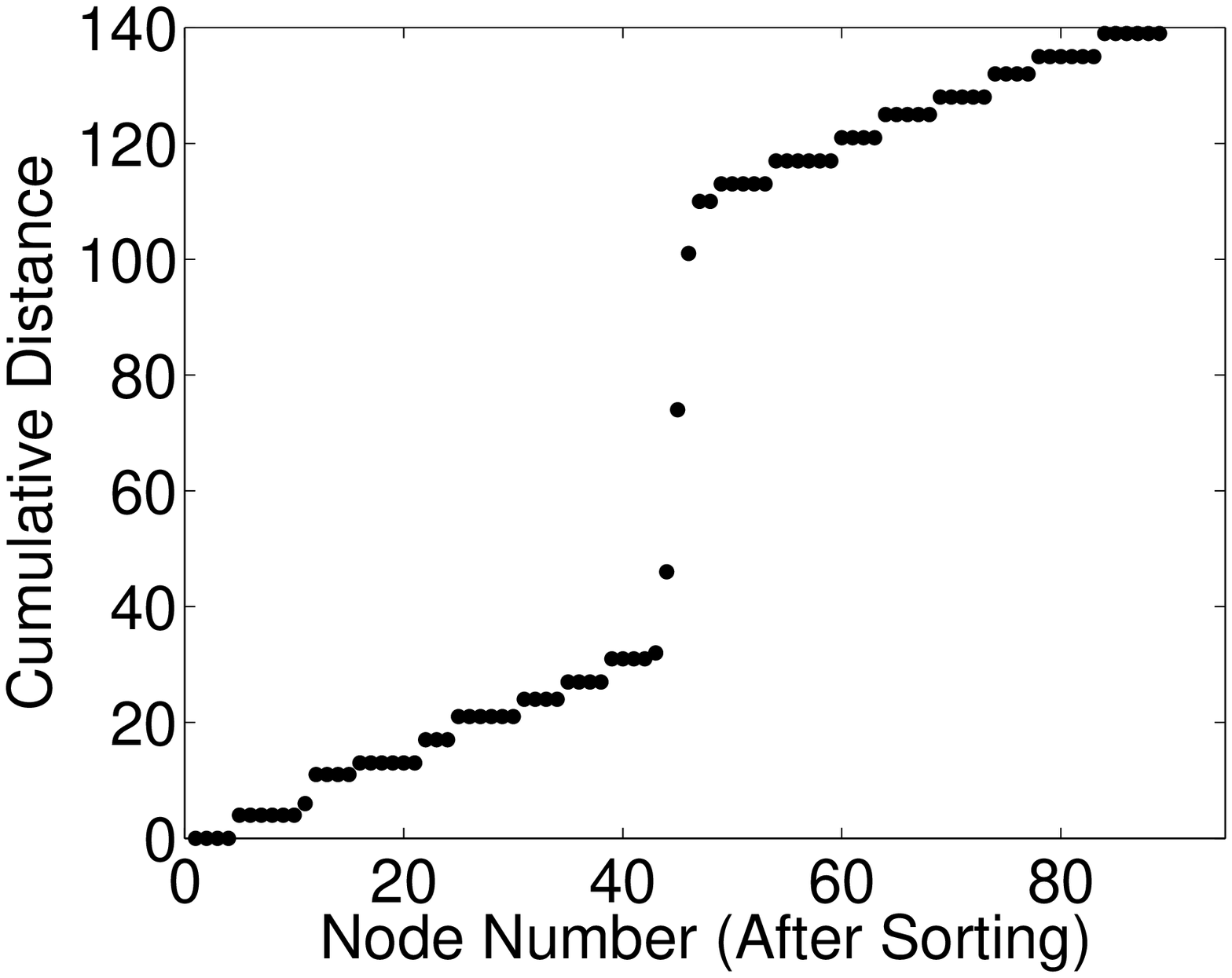}
				\caption{Row Distances for Ideal graph 3, $\alpha=0.87$.  Generated using the membership matrix shown in Figure \ref{Ideal3MM}}
				\label{Ideal3RD}
			\end{center}
		\end{minipage}
		\hspace{0.5cm} 
		\begin{minipage}[b]{0.475\linewidth} 
			\begin{center}
				\includegraphics[width=\textwidth]{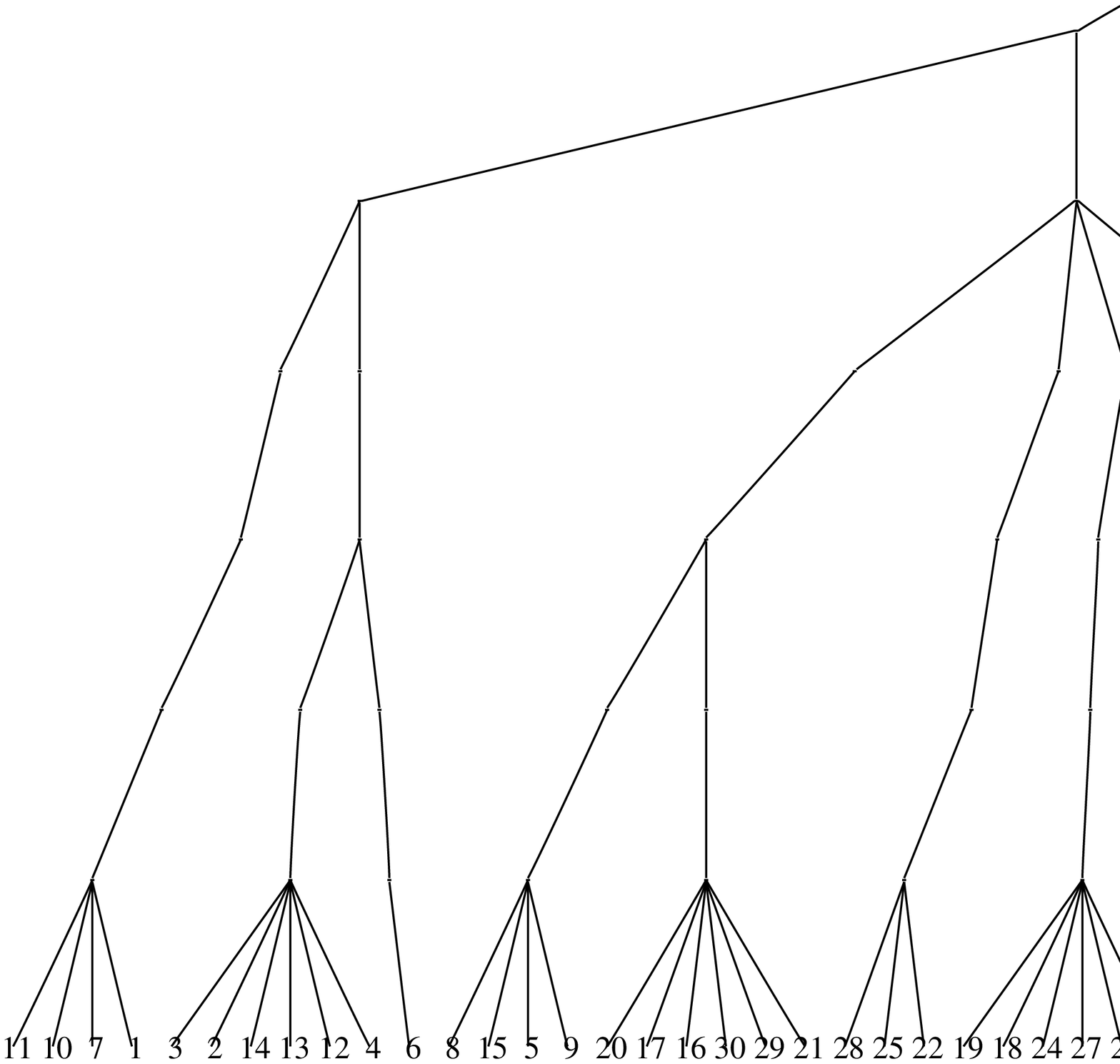}
				\caption{Dendrogram for Ideal graph 2, $\alpha=0.87$.  Now with central members, for different scale communities, the dendrogram becomes more complicated in contrast to those in Figures \ref{Ideal1Dendo} and \ref{Ideal2Dendo}.}
				\label{Ideal3Dendo}
			\end{center}
		\end{minipage}
	\end{figure}
	
\end{section}

\begin{section}{Real-World Networks}\label{RealWorldNets}
	The proposed algorithm performs extremely well on idealized networks (see Section \ref{Motivating}), but how does it perform on real-world networks? Here we analyze the Zachary Karate Club, the network of co-appearances present in the novel Les Miserables by Victor Hugo, and the partisan network of co-purchased books on American politics.

	\begin{subsection}{Zachary Karate Club}
		The Zachary Karate Club is perhaps the most famous network in terms of community structure~\cite{ZacharyKarate}.  The club suffered from infighting and eventually split in half, providing actual evidence of the community structure, at least at the top-most level.  Thus, it provides an excellent means to compare the accuracy of any proposed detection methods.

		For $\alpha=1.2$, we achieve a reasonable result:  three vertices (3, 14, and 20) are labeled incorrectly as compared with the betweenness partitioning~\cite{NewmanLesMis} and the actual split the club underwent~\cite{ZacharyKarate}.  Looking at the network itself (Figure \ref{KarateGRAPH}), all of the disputed vertices are situated on the ``border" of one community or the other.  One would expect these vertices to be the most likely to be labeled incorrectly because, unlike more idealized networks, these vertices are almost equally connected to both communities.

		Figures \ref{KarateMMafterSort}, \ref{KarateRD}, and \ref{KarateDendo} contain the membership matrix, cumulative row distances plot, 	and dendrogram, respectively. 

		\begin{figure}
			\begin{minipage}[b]{0.475\linewidth} 
				\begin{center}
					\includegraphics[width=0.65\textwidth]{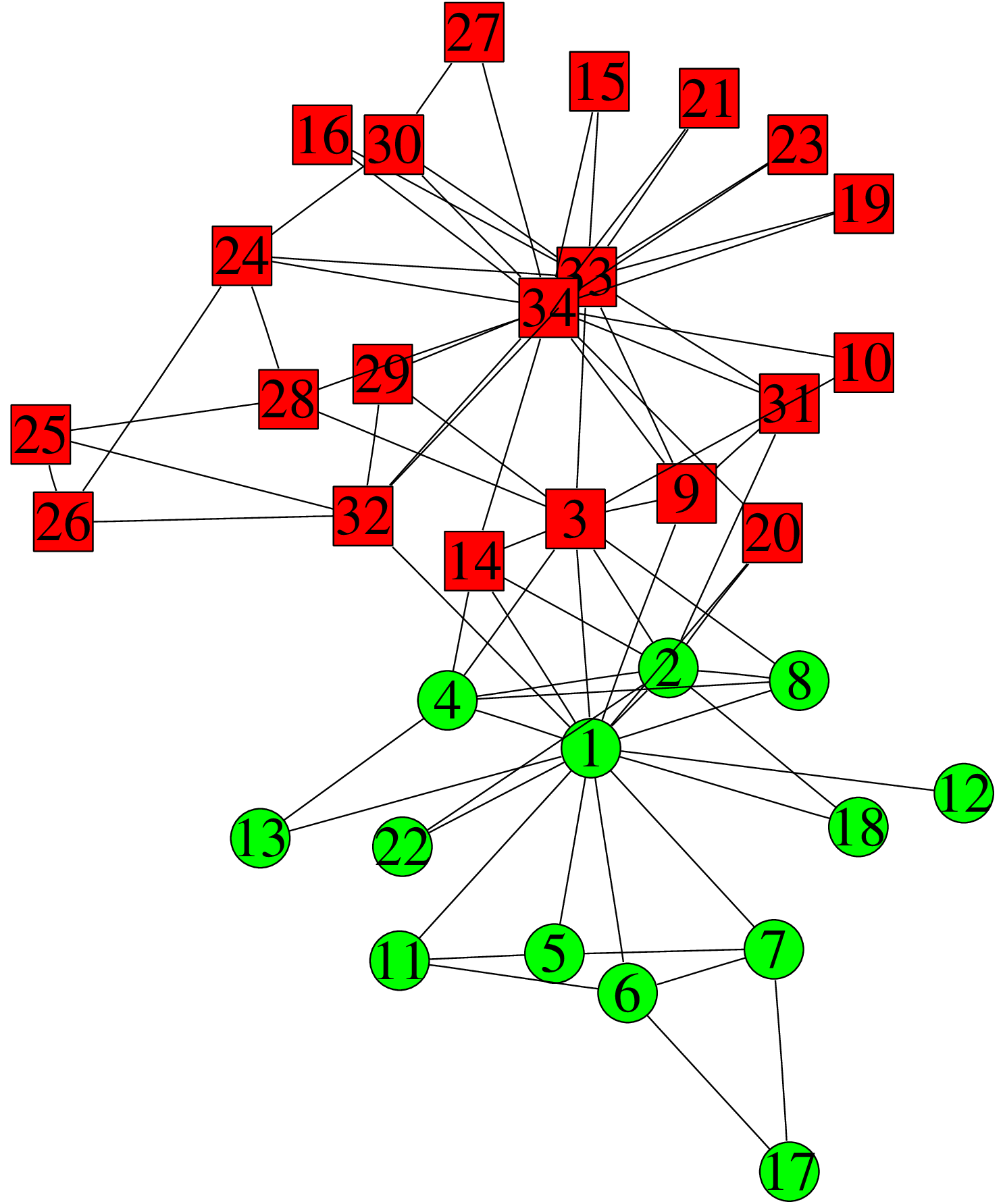}
					\caption{The Zachary Karate club.  The coloring indicates the membership of the two clusters of the topmost branch of 	the dendrogram (Figure \protect\ref{KarateDendo})}
					\label{KarateGRAPH}
				\end{center}
			\end{minipage}
			\hspace{0.5cm} 
			\begin{minipage}[b]{0.475\linewidth}
				\begin{center}
					\includegraphics[width=\textwidth]{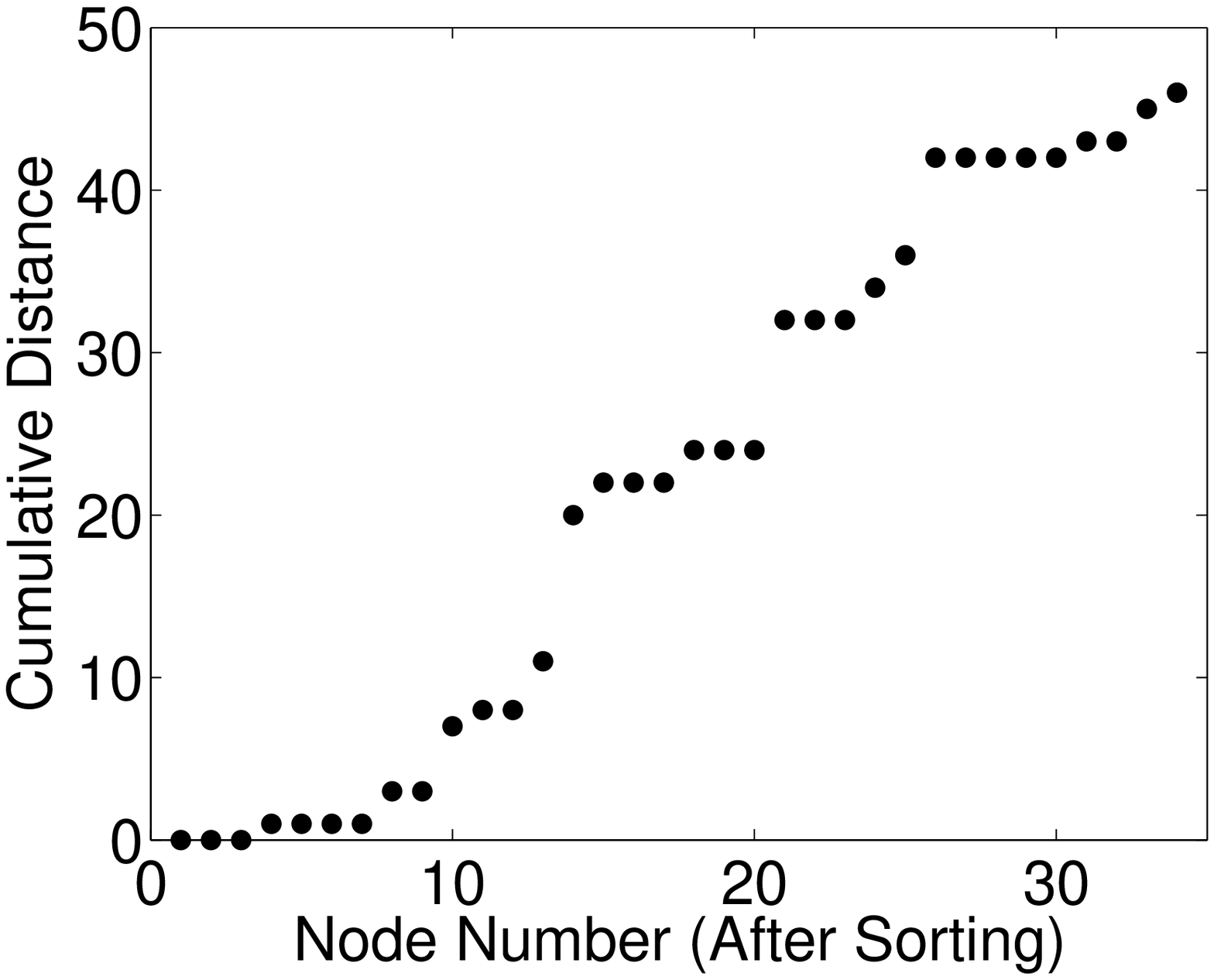}
					\caption{Cumulative row distances for the Karate club, computed using the membership matrix shown in Figure 								\protect\ref{KarateMMafterSort} ($\alpha = 1.2$). }
					\label{KarateRD}
				\end{center}
			\end{minipage}
		\end{figure}
		\begin{figure}
			\begin{center}
				\includegraphics[width=\textwidth]{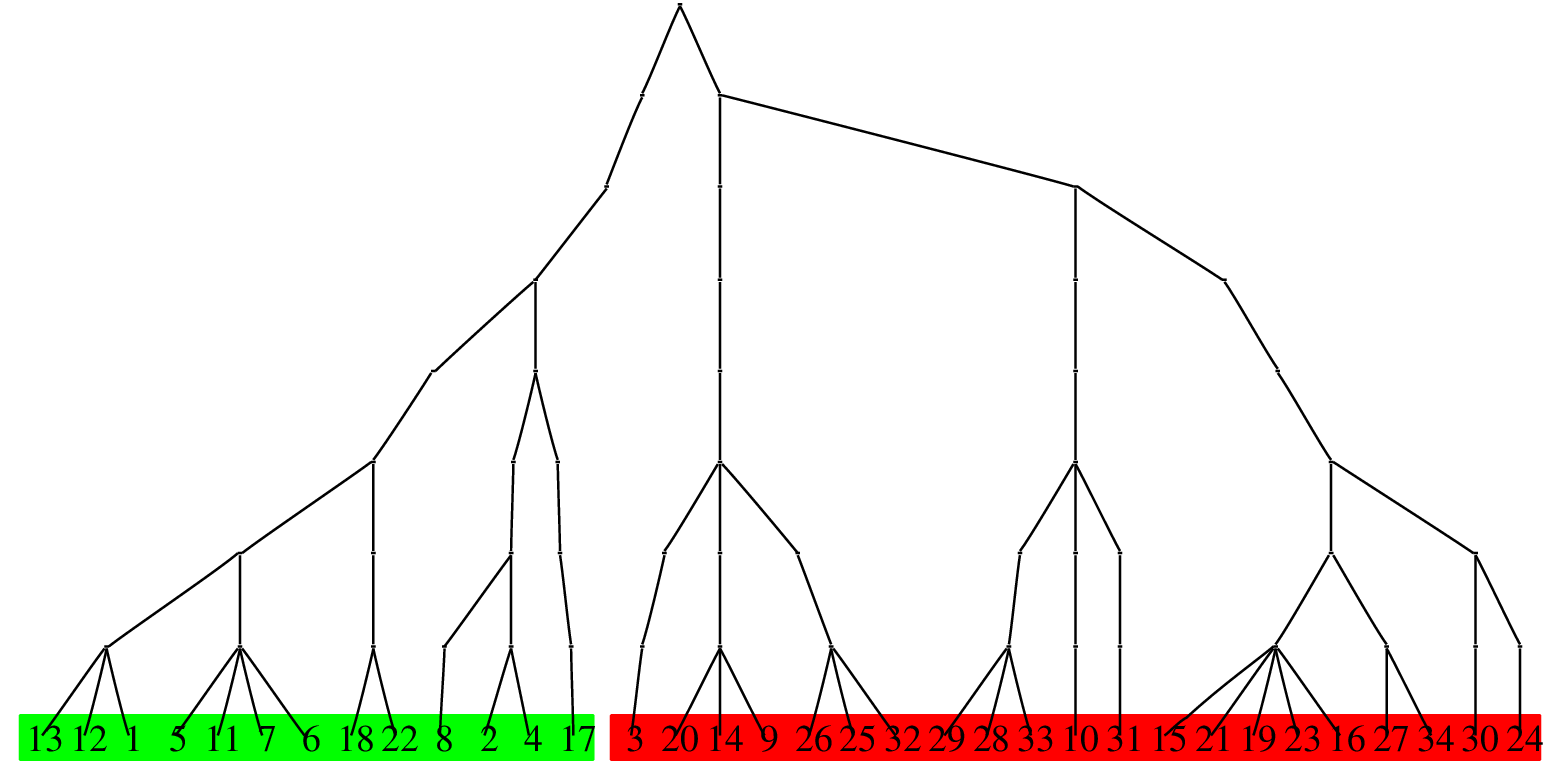}
				\caption{Dendrogram for the Karate club, using the membership matrix shown in Figure \protect\ref{KarateMMafterSort} ($\alpha = 1.2$).}
				\label{KarateDendo}
			\end{center}
		\end{figure}
	\end{subsection}
	
	\begin{subsection}{Books on Politics}
		A network possessing a fairly simple two-community structure is the network of co-purchased books on American politics  shown in Figure \ref{BooksGraph}~\cite{BooksPolitics}.  As can be seen from Figures \ref{BooksMM} and \ref{BooksDendo}, the results are extremely reasonable.  
		
		\begin{figure}
			\begin{minipage}[t]{0.475\linewidth} 
				\begin{center}
					\includegraphics[width=.6\textwidth]{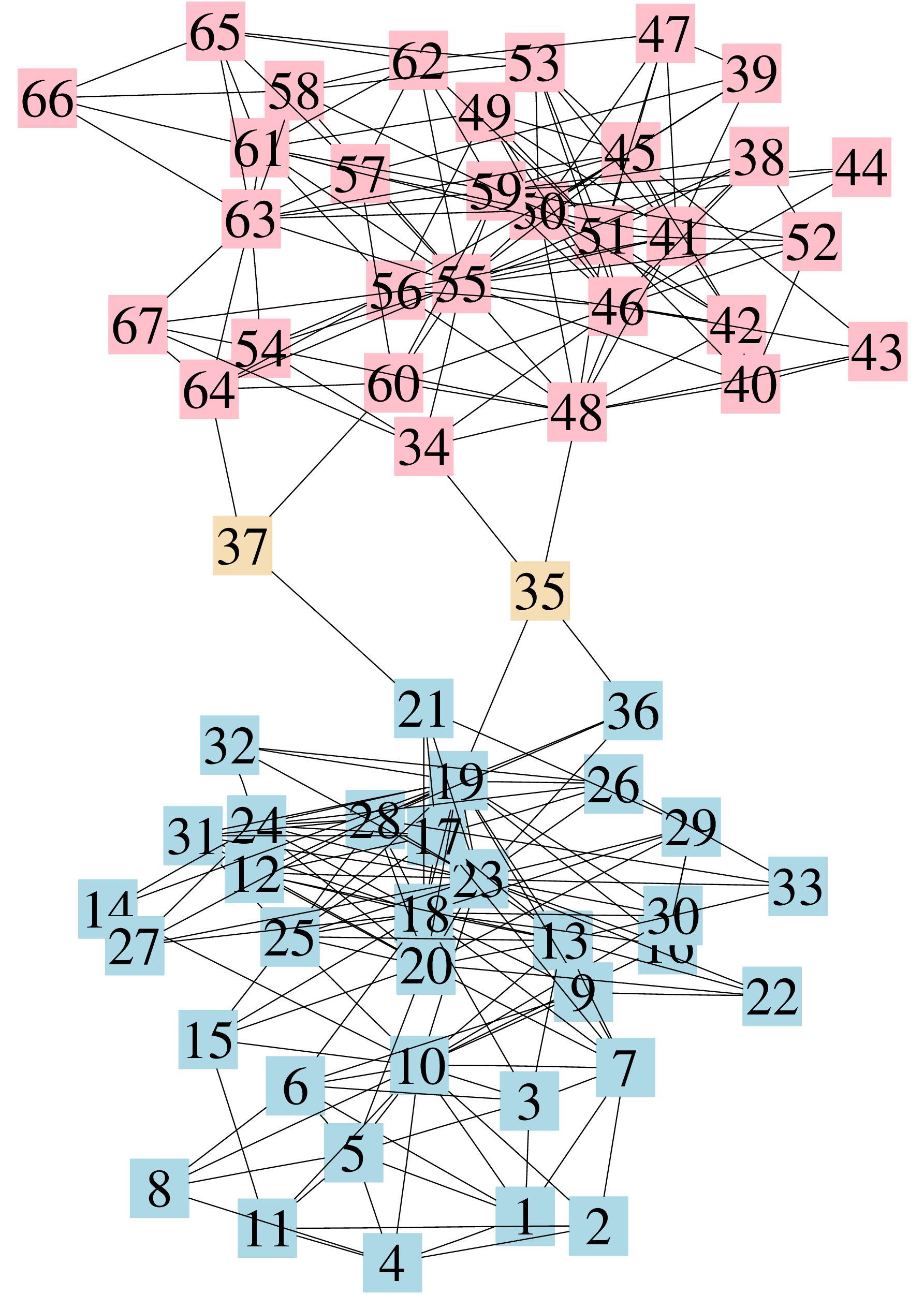}
					\caption{The network of co-purchased books on American politics~\protect\cite{BooksPolitics}.  Here a link is drawn between two vertices if those books were purchased together from a major online retailer.}
					\label{BooksGraph}
				\end{center}
			\end{minipage}
			\hspace{0.5cm} 
			\begin{minipage}[t]{0.475\linewidth}
				\begin{center}
					\includegraphics[width=\textwidth]{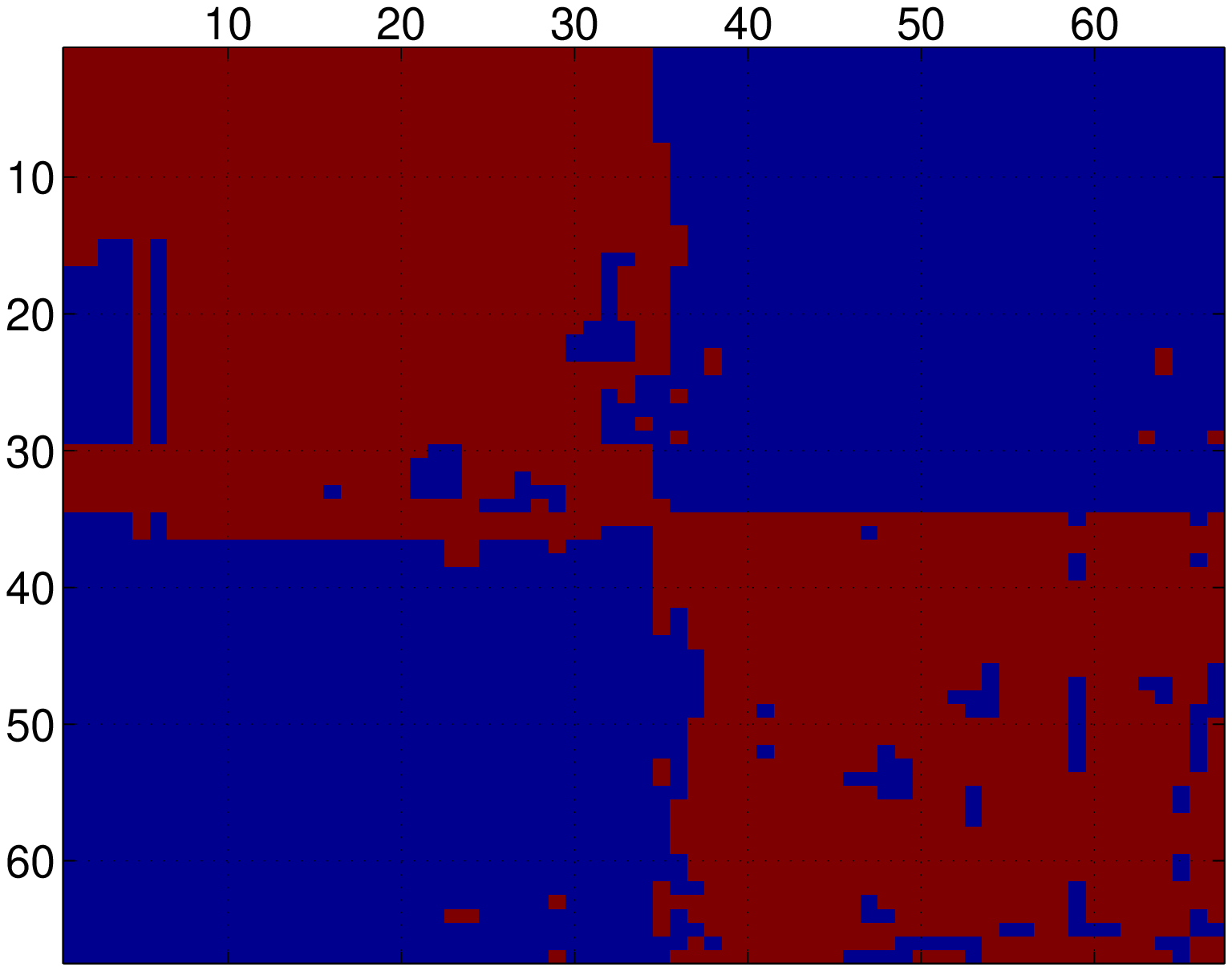}
					\caption{Membership matrix  for the Books on Politics network. $\alpha=1.2$.}
					\label{BooksMM}
				\end{center}
			\end{minipage}
		\end{figure}
		\begin{figure}[!h]
			\begin{center}
				\includegraphics[width=\textwidth]{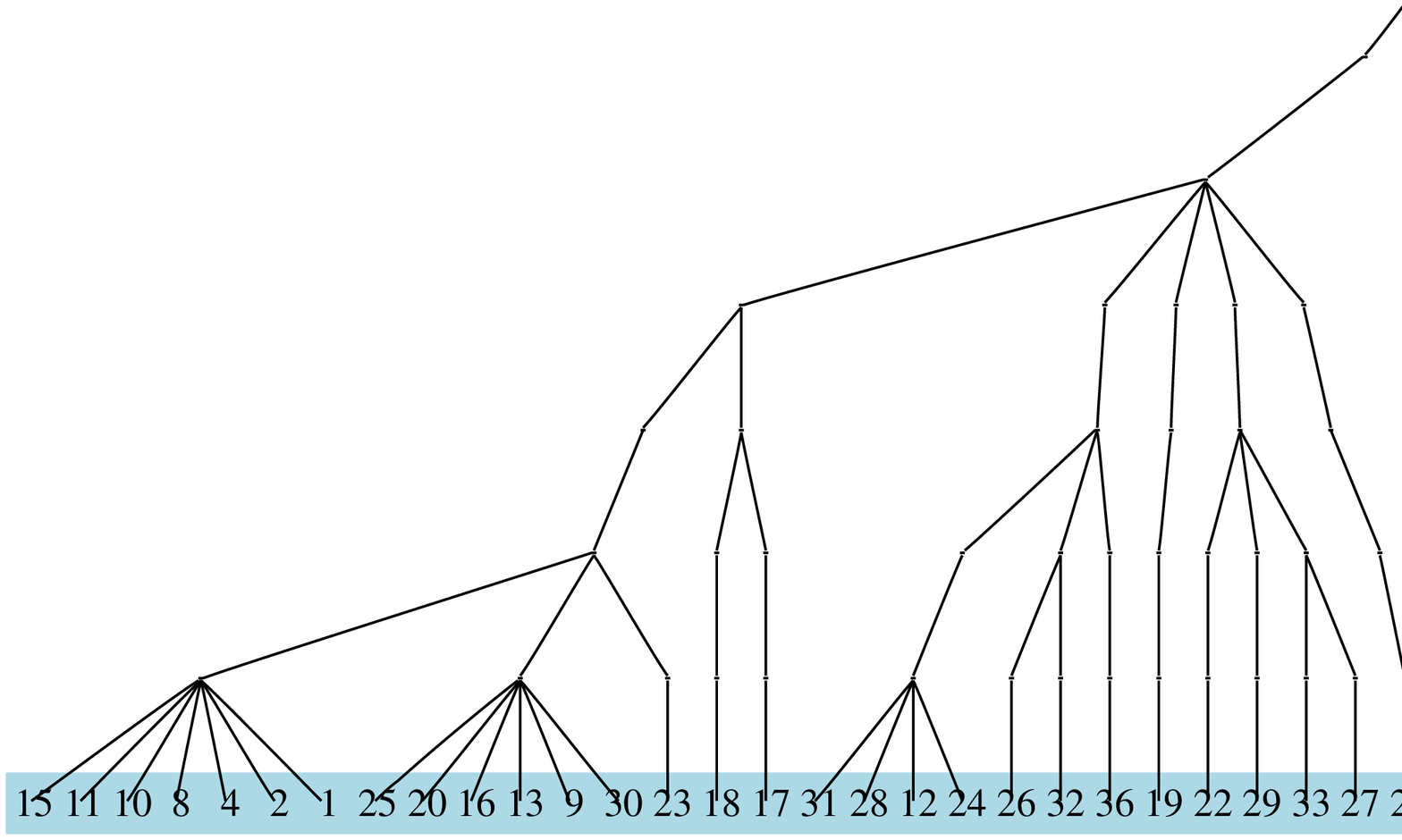}
				\caption{Dendrogram for the Books on Politics network.  Generated using the membership matrix shown in Figure \protect\ref{BooksMM}.  Note that vertices 35 and 37 are rows 34 and 35, respectively, of the membership matrix.}
				\label{BooksDendo}
			\end{center}
		\end{figure}
	\end{subsection}
	
	\begin{subsection}{Les Miserables}
		Another network with an interesting community structure is the network of character co-appearances from the novel Les Miserables by Victor Hugo \cite{NewmanLesMis}.  This network, shown in Figure \ref{LesMisGraph}, differs from the karate club and the political books networks in that there are several communities of smaller size rather than two large communities.  
		As can be seen from the membership matrix in Figure \ref{LesMisMM}, some of the communities separated quite well, while others were detected rather poorly, at least compared to the results in \cite{NewmanLesMis}. 		
			
		\begin{figure}
			\begin{minipage}[t]{0.475\linewidth} 
				\begin{center}
					\includegraphics[width=.8\textwidth]{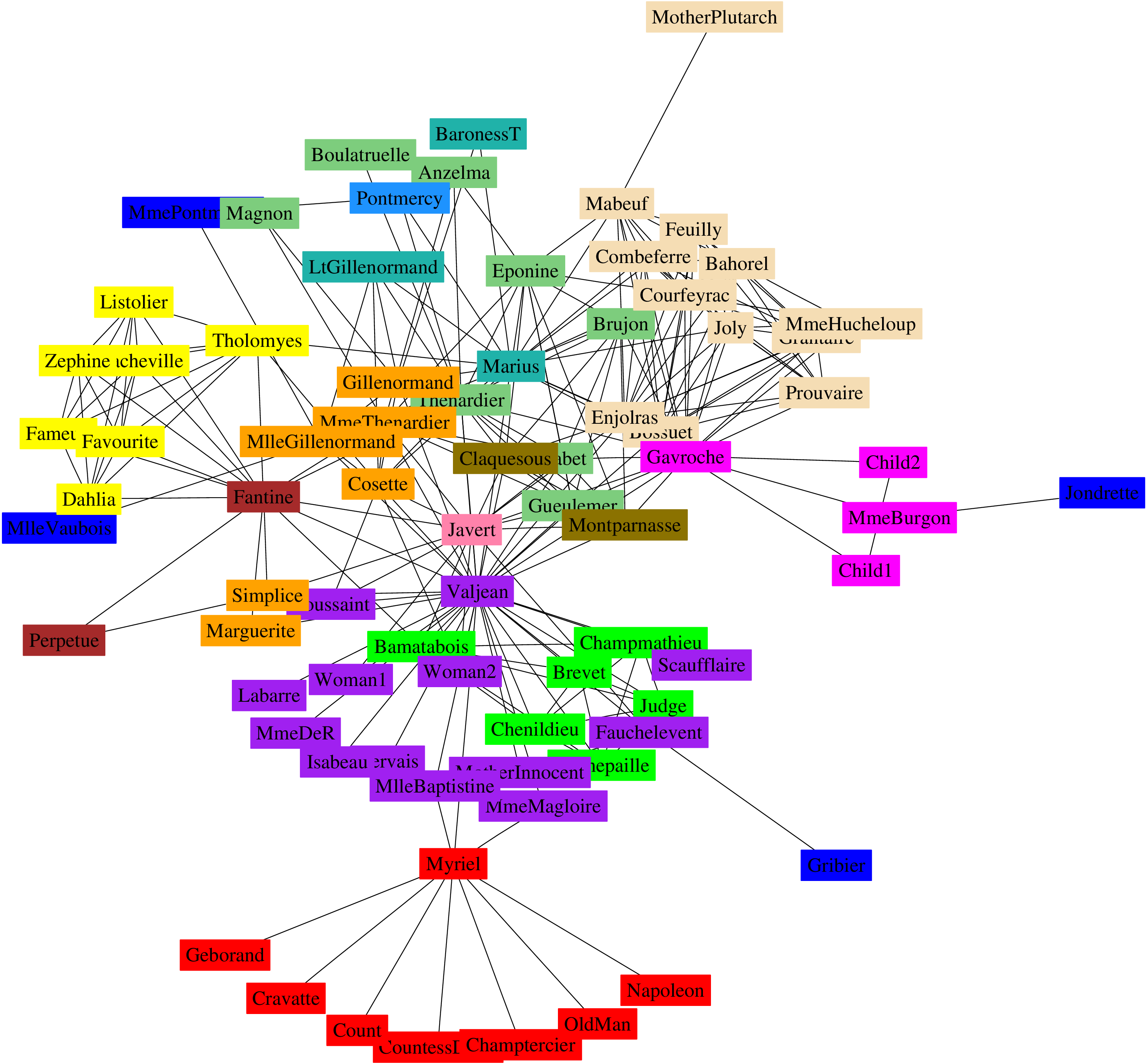}
					\caption{The network of  character co-appearances from the novel Les Miserables by Victor Hugo. See Figure \protect\ref{LesMisDendo} for partitioning.}
					\label{LesMisGraph}
				\end{center}
			\end{minipage}
			\hspace{0.5cm} 
			\begin{minipage}[t]{0.475\linewidth}
				\begin{center}
					\includegraphics[width=\textwidth]{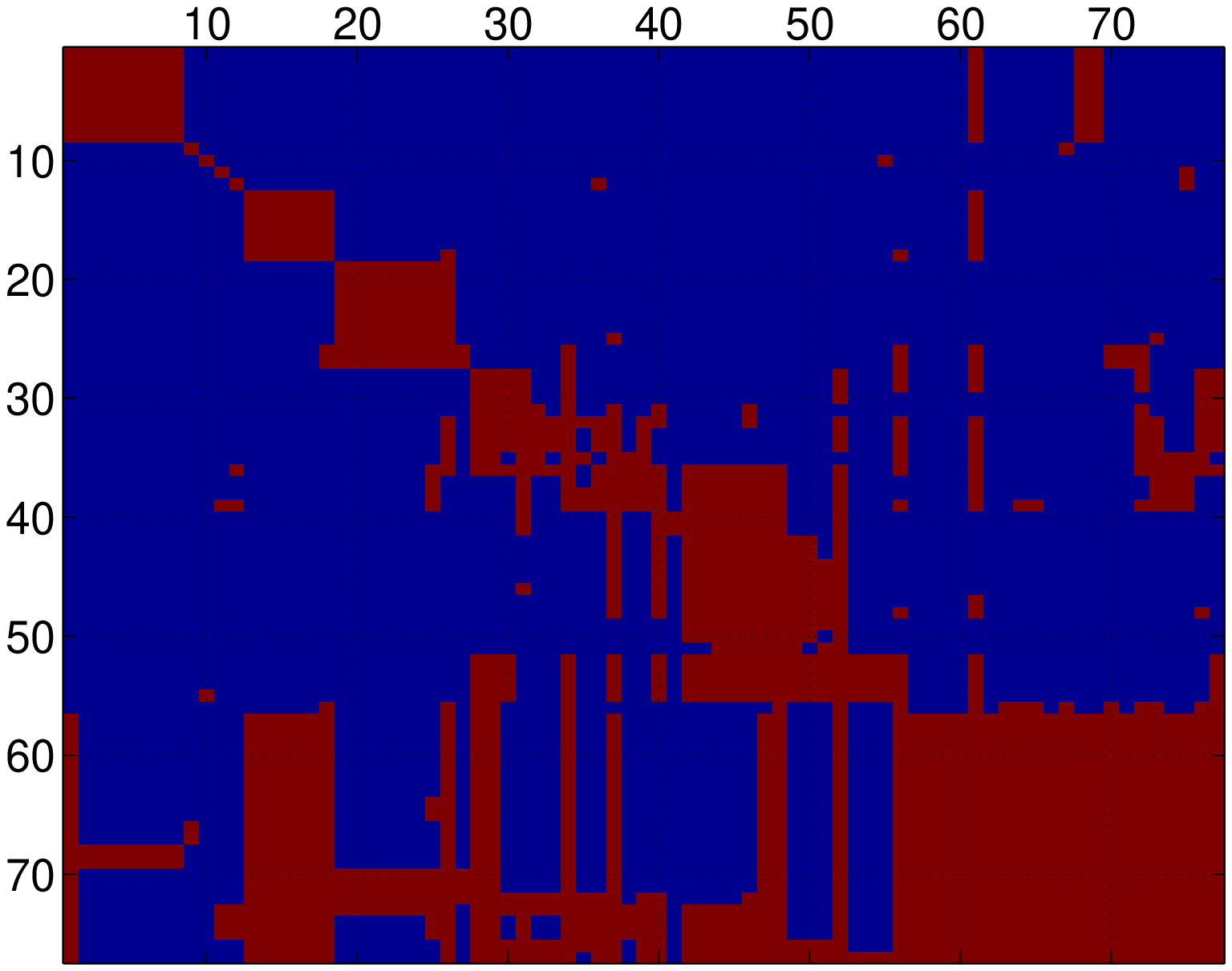}
					\caption{Membership matrix (after sorting) for the Les Mis network. $\alpha=6.9$.}
					\label{LesMisMM}
				\end{center}
			\end{minipage}
		\end{figure}

		\begin{figure}
			\begin{center}
				\includegraphics[width=\textwidth]{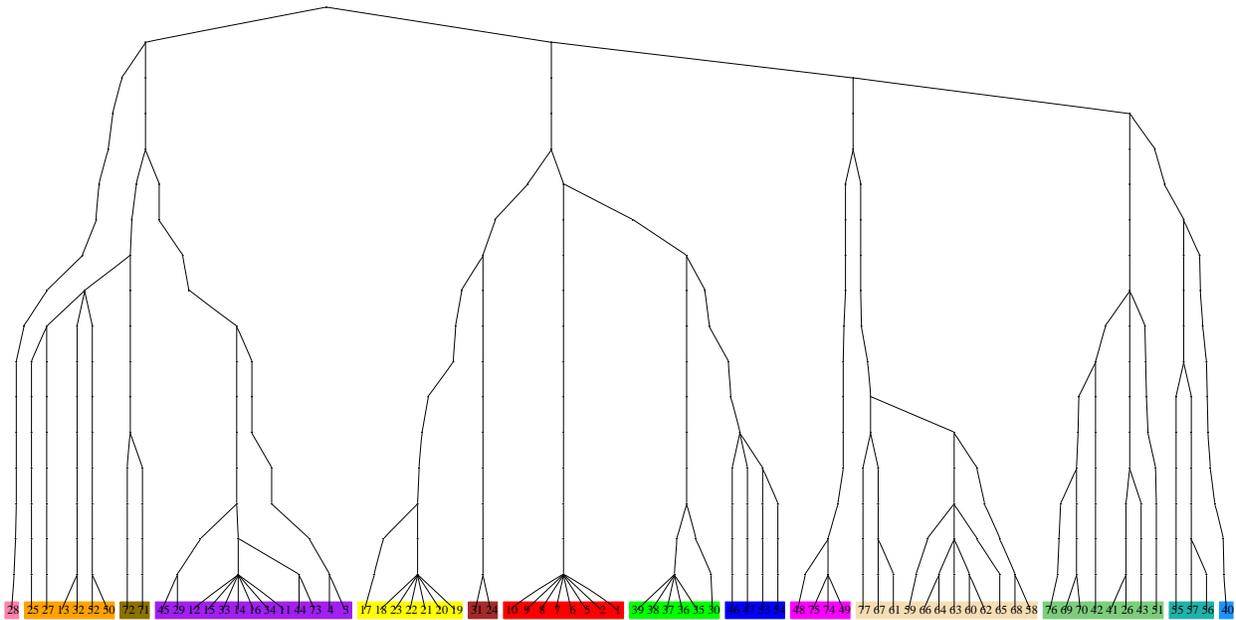}
				\caption{Dendrogram for the Les Mis social network.  Generated using the membership matrix shown in Figure \protect\ref{LesMisMM}. }
				\label{LesMisDendo}
			\end{center}
		\end{figure}
	\end{subsection} 
\end{section}

\begin{section}{Conclusions}
	In this paper, we have introduced Algorithm \ref{localAlg.alg}, a new method for detecting community structure.  This method is local and may be applied in situations where other methods are too inefficient; for example, when one is concerned with a single community and not the complete community structure of a graph.  A single parameter, $\alpha$, is used, making it very easy to tune the output of the algorithm, as was shown in Section \ref{ImpactAlpha}.  
	
	We have also proposed one possible method of applying Algorithm \ref{localAlg.alg} globally (see Sections \ref{globalinfo} and \ref{SubComms}).  Sorting the membership matrix is expensive and limits  this global application to smaller networks. In addition, the membership matrix is, unlike an adjacency matrix, not necessarily sparse for a sparse graph: it will only be sparse when the sizes of the individual communities are all much smaller than the size of the network as a whole.  This limits the possibility of replacing the membership matrix with a more efficient data structure.
	
	
	We feel that Algorithm \ref{localAlg.alg} is a useful result, due to both its simplicity and flexibility.  For example, our global application could be easily altered to become more efficient: one imagines the cost of the sorting algorithm can be offset a great deal by only starting Algorithm \ref{localAlg.alg} from a fraction of the vertices, $f$, instead of all vertices in the graph.  This reduces the cost of the sorting algorithm by a factor of $f^3$, a real savings for small $f$, with the presumed tradeoff being a reduction in accuracy as $f$ decreases.   Other applications of Algorithm \ref{localAlg.alg} besides the expensive use of the membership matrix may also be discovered.
	
	The Zachary Karate club has become an almost canonical representative of a community structure.  The possibility remains, however, that outside factors may not be represented in the dataset.  If this is true, then the club's fissure should not be used as the sole means of justification for a community detection method.  For example, some of the border nodes, represented as diamonds in Figure \ref{KarateLocals}, could well have joined either community, based solely on the network at hand.   This can lead to ambiguity in the final partition.  The point is that the algorithm defines the community; the community should not define the algorithm.     
	
	Another concept, often neglected in determining communities, is the idea of a relative result.  Who is to say that someone in a town agrees with what community the rest of the town feels he or she belongs to?  It seems feasible that a vertex that is equally linked to two communities in a graph is just as likely to correspond to a person who thinks he or she is a member of both communities as it is to correspond to a person who feels they are independent of both larger communities.   If one considers the output of Algorithm \ref{localAlg.alg} to be what the starting vertex ``believes" to be his or her community, then this method may prove to be a useful tool when analyzing relative results.  This is not useful for many applications of community detection where a final structure is necessary, such as minimizing crosstalk between parallel processors in a computer, but it may prove very useful in areas such as social networks.  
	  
\end{section}

\begin{section}*{Acknowledgements}
	The authors are grateful to Hernan Rozenfeld and Daniel ben-Avraham for useful discussions and to Mark Newman for useful data. This work was supported by the NSF under grants DMS0404778.
\end{section}

\end{document}